\documentclass{aa}
\usepackage{graphicx}
\usepackage{amsmath,amssymb}
\usepackage[applemac]{inputenc}
\usepackage[modulo, switch]{lineno}
\usepackage{xcolor}
\usepackage{colortbl}
\usepackage{txfonts}
\usepackage{natbib}
\bibpunct{(}{)}{;}{a}{}{,} % to follow the A&A style
\newcommand{\kepler}{\textit{Kepler} }

\newcommand{\Dnu}{\Delta \nu}
\newcommand{\numax}{\nu_\mathrm{max}}

\newcommand{\muhz}{$\mu$Hz}

\newcommand{\diamonds}{\textsc{D\large{iamonds}}}
\newcommand{\ngca}{NGC\,6791}
\newcommand{\ngcb}{NGC\,6819}
\newcommand{\ngcc}{NGC\,6811}

\def\be{\begin{equation}}
\def\ee{\end{equation}}    
\def\ba{\begin{eqnarray}}
\def\ea{\end{eqnarray}}

\begin{document}

\title{Metallicity effect on stellar granulation detected from oscillating red giants in open clusters}
\author{
E.~Corsaro\inst{1,2,3,4},
S.~Mathur\inst{5},
R.~A.~Garc\'{i}a\inst{1},
P.~Gaulme\inst{6,7,8},
M.~Pinsonneault\inst{9},
K.~Stassun\inst{10},
D.~Stello\inst{11,12,13},
J.~Tayar\inst{9},
R.~Trampedach\inst{5,13},
C.~Jiang\inst{14},
C.~Nitschelm\inst{15},
D.~Salabert\inst{4}
}

\offprints{Enrico Corsaro\\ \email{enrico.corsaro@oact.inaf.it}}

\institute{INAF - Osservatorio Astrofisico di Catania, Via S. Sofia 78, I-95123 Catania, Italy
\and Instituto de Astrof\'{i}sica de Canarias, E-38205, Tenerife, Spain
\and Universidad de La Laguna, Departamento de Astrof\'{i}sica, E-38206, La Laguna, Tenerife, Spain
\and Laboratoire AIM, CEA/DRF -- CNRS -- Univ. Paris Diderot -- IRFU/SAp, Centre de Saclay, 91191 Gif-sur-Yvette Cedex, France
\and Space Science Institute, 4750 Walnut street Suite\#205, Boulder CO 80301, USA
\and Department of Astronomy, New Mexico State University, P.O. Box 30001, MSC 4500, Las Cruces, NM 88003-8001, USA
\and Apache Point Observatory, 2001 Apache Point Road, P.O. Box 59, Sunspot, NM 88349, USA
\and Physics Department, New Mexico Institute of Mining and Technology, 801 Leroy Place, Socorro, NM 87801, USA
\and Department of Astronomy, Ohio State University, 140 W 18th Ave, OH 43210, USA
\and Department of Physics and Astronomy, Vanderbilt University, 1807 Station B, Nashville, Tennessee 37235, USA
\and School of Physics, University of New South Wales, NSW, 2052, Australia
\and Sydney Institute for Astronomy (SIfA), School of Physics, University of Sydney, NSW 2006, Australia
\and Stellar Astrophysics Centre, Department of Physics and Astronomy, Aarhus University, Ny Munkegade 120, DK-8000 Aarhus C, Denmark
\and Instituto de Astrof\'{i}sica e Ci\^{e}ncias do Espa\c{c}o, Universidade do Porto, CAUP, Rua das Estrelas, 4150-762 Porto, Portugal
\and Unidad de Astronom\'{i}a, Universidad de Antofagasta, Avenida Angamos 601, Antofagasta 1270300, Chile
}

   \date{Received 3 May 2017; Accepted 19 July 2017}

\abstract
{The effect of metallicity on the granulation activity in stars, and hence on the convective motions in general, is still poorly understood. Available spectroscopic parameters from the updated APOGEE-\kepler catalog, coupled with high-precision photometric observations from NASA's \kepler mission spanning more than four years of observation, make oscillating red giant stars in open clusters crucial testbeds.}
{We aim to determine the role of metallicity on the stellar granulation activity by discriminating its effect from that of different stellar properties such as surface gravity, mass, and temperature. We analyze 60 known red giant stars belonging to the open clusters NGC 6791, NGC 6819, and NGC 6811, spanning a metallicity range from [Fe/H] $\simeq -0.09$ to $0.32$. The parameters describing the granulation activity of these stars and their frequency of maximum oscillation power, $\numax$, are studied while taking into account different masses, metallicities, and stellar evolutionary stages. We derive new scaling relations for the granulation activity, re-calibrate existing ones, and identify the best scaling relations from the available set of observations.}
{We adopted the Bayesian code \diamonds\,for the analysis of the background signal in the Fourier spectra of the stars. We performed a Bayesian parameter estimation and model comparison to test the different model hypotheses proposed in this work and in the literature.}
{Metallicity causes a statistically significant change in the amplitude of the granulation activity, with a dependency stronger than that induced by both stellar mass and surface gravity. We also find that the metallicity has a significant impact on the corresponding time scales of the phenomenon. The effect of metallicity on the time scale is stronger than that of mass.}
{A higher metallicity increases the amplitude of granulation and meso-granulation signals and slows down their characteristic time scales toward longer periods. The trend in amplitude is in qualitative agreement with predictions from existing 3D hydrodynamical simulations of stellar atmospheres from main sequence to red giant stars. We confirm that the granulation activity is not sensitive to changes in the stellar core and that it only depends on the atmospheric parameters of stars.}

% 5 {} token are mandatory
%
%\abstract
%{ }
\titlerunning{Metallicity effect on stellar granulation from cluster red giants}
      \authorrunning{E. Corsaro et al.}
\maketitle
%
%==========================================================================
\section{Introduction}
\label{sec:intro}
Granulation is a type of stellar variability and it is a surface manifestation of stellar envelope convection. Here hot gas in the granules rises from the interior to the photosphere where the thermal energy of the granules is lost to the radiation field, reaching velocities comparable to the local sound speed. The cooled, denser plasma is thus pushed to the edges of the granules and sinks back into the star in the darker inter-granular lanes. According to this interpretation, a characteristic time scale for the phenomenon is to first approximation given as $\propto \sqrt{T_\mathrm{eff}} / g$ \citep{Brown91numax,Kjeldsen11}, where $g$ is the surface gravity of the star. For solar-like oscillating stars, acoustic oscillations also originate from the turbulent motions caused by convection, although granulation remains the dominant component in terms of energy that is visible at the stellar surface because the intensity fluctuation related to granulation can be up to about three times that related to the acoustic oscillation signal \citep[][hereafter K14]{Kallinger14}. 

The study of stellar granulation was born through its observation on the Sun \citep{Herschel1801}. The first analysis using the Fourier approach to measure the granulation time scale and amplitude was done by \cite{Harvey1985}, and subsequently improved by for example, \cite{Aigrain04bkg}. Since then, granulation activity has been observed in a large variety and number of low- and intermediate-mass stars with convective envelopes \citep[e.g.,][K14]{Kallinger10gran,Mathur11granulation,Hekker12variance,Karoff13}. It has also been used to obtain accurate model-independent measurements of stellar surface gravity \citep{Bastien13Nature,Bastien16,Kallinger16Science}. These studies have been made possible thanks to the advent of high-precision photometry from space missions such as CoRoT \citep{Baglin06} and NASA \kepler \citep{Borucki10}, the latter having been used to observe more than 197,000 stars \citep{Mathur17}. These space missions provided both sampling rates rapid enough for resolving the typical time scales of granulation, and observing lengths that allowed for characterization of the granulation properties to a high degree of precision and accuracy. In particular, the first ensemble study was done for red giants (RGs) by \cite{Mathur11granulation} using \kepler data spanning more than one year of nearly-continuous observations. The authors showed that the granulation power and time scale are strongly correlated with the frequency of maximum oscillation power, $\numax$, the latter scaling with the acoustic cut-off frequency of the star \citep{Brown91numax}. Later K14 provided a thorough calibration of these dependencies by extending the sample to main sequence stars and using \kepler observations covering more than three years.

Studying the connection between the granulation signal and fundamental stellar properties such as surface gravity, mass, temperature, and chemical composition is essential to better understand convection in stars. A better understanding of stellar granulation can yield more detailed descriptions of turbulent motions in stellar atmospheres, and therefore improve stellar structure and evolution models. More realistic stellar models improve our capability to retrieve accurate stellar properties, and provide high-quality evolution sequences for ensemble analysis of, for example, the Galactic formation and evolution, especially in view of the ESA Gaia mission \citep{Perryman01Gaia}. Efforts in this direction have been made from a theoretical point of view by using 3D hydrodynamical models of stellar atmospheres \cite[e.g.,][]{Trampedach98,Ludwig06gran,Mathur11granulation,Samadi13a,Samadi13b,Trampedach13gran,Trampedach14}, although only a few studies \citep[e.g.,][]{Collet07granulation,Magic15a,Magic15b} have dealt with metallicity effects on such 3D simulations of convective atmospheres. As shown by \cite{Collet07granulation} for RGs \citep[see also the work by][on gray atmospheres of main sequence stars]{Tanner13}, stellar metallicity appears to play an important role in determining the scale of granulation, yielding larger granules as metallicity increases, hence a higher amplitude of the associated granulation signal \citep{Ludwig06gran}. This result has been further confirmed for evolved stars by \cite{Ludwig16}. However, any observational evidence of the metallicity effect on granulation has neither been found nor discussed in the literature until now.

Stellar clusters offer a possibility to exploit the accurate knowledge of the common physical properties shared by their members. The open clusters NGC\,6791, NGC\,6819, and NGC\,6811 have been monitored by the \kepler mission for more than four years, thus providing us with the best photometric observations currently available for the rich populations of RGs hosted by each of these clusters \citep{Stello11membership}. Fundamental parameters such as temperature, mass, metallicity, and age, are determined for cluster stars with high reliability \citep[e.g.,][]{Bragaglia01OC,Basu11OC,Hekker11OC,Stello11membership,Brogaard11,Miglio12massloss,Brogaard12}, and the evolutionary stage of many cluster RGs is also well known from existing asteroseismic analyses \citep{Corsaro12,Mosser14,Vrard16,Corsaro17}. 

In this work we have exploited the full \kepler nominal mission photometric data for the open clusters NGC\,6791, NGC\,6819 and NGC\,6811, and the wealth of spectroscopic observations available from APOKASC \citep{Pinsonneault14APOKASC}, to properly disentangle the effect of metallicity from that of other fundamental stellar properties by performing a thorough Bayesian approach that takes into account uncertainties on all the observables. In this way we will assess the behavior of granulation activity in RGs in light of existing theoretical predictions.

\section{Observations and data} 
\label{sec:data}

\subsection{Sample selection and photometry}
\label{sec:photometry}
The sample of RGs of the open clusters \ngca\, and \ngcb\, is derived from the original set of 111 stars analyzed by \cite{Corsaro12}. We included those stars with a clear evolutionary stage determination, as discriminated using mixed mode oscillations \citep{Bedding11Nature} by \cite{Corsaro12,Mosser14,Corsaro17} (see also Corsaro et al. in prep.). We find in total 30 RGs for \ngca\,and 24 for \ngcb. For \ngcc, we considered the four stars with a known evolutionary stage from \cite{Corsaro12} and we added two more, KIC~9776739 and KIC~9716090, analyzed by \cite{Molenda14NGC6811} and by Corsaro et al. (in prep.) and both classified as core-He-burning RGs (red clump stars, hereafter RC), thus reaching a total of six targets for this cluster. The final sample therefore accounts for 60 RGs, with 38 RC stars and 22 shell-H-burning RGs (red giant branch, hereafter RGB). 

\begin{figure}
   \centering
   \includegraphics[width=9.0cm]{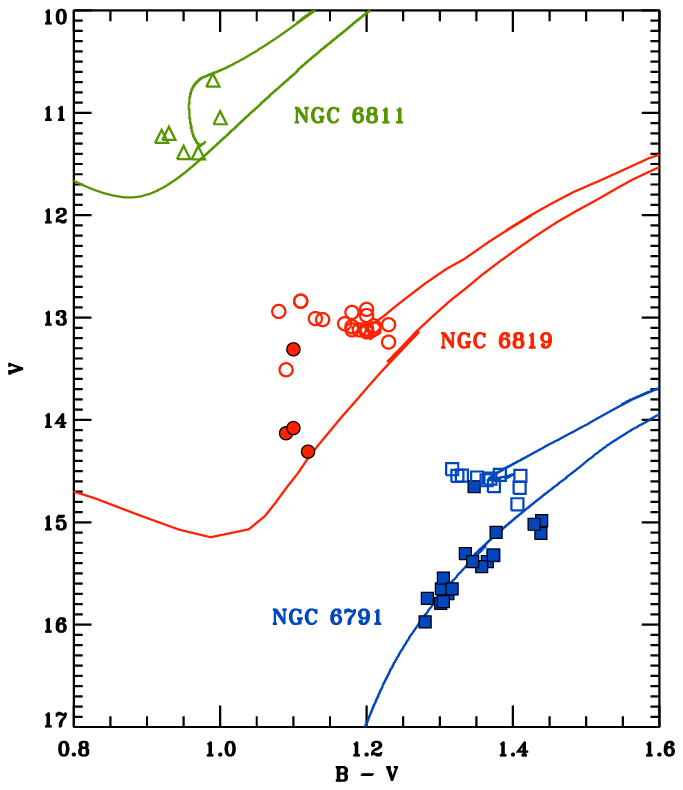}
      \caption{Color-magnitude diagram of the three open clusters NGC\,6791 (blue squares), NGC\,6819 (red circles), and NGC\,6811 (green triangles), with color and magnitudes of the 60 cluster RGs sourced from \cite{Stello11membership,Corsaro12}. Hydrogen-shell-burning and core-He-burning RGs are shown with filled and open symbols, respectively, with an evolutionary stage identified according to \cite{Corsaro12,Corsaro17}, and to \cite{Molenda14NGC6811} for the two stars KIC~9776739 and KIC~9716090. Isochrones are shown for each cluster as solid lines (see \citealt{Stello11membership} for more details).}
    \label{fig:cmd}
\end{figure}

The photometric observations for the selected sample of stars were conducted by NASA's \kepler telescope in the observing quarters (Q) from Q0 till Q17, for a total of $\sim 1460$ days in long cadence mode \citep{Jenkins10}. All the original light curves were processed and optimized for asteroseismic analysis following \cite{Garcia11,Garcia14}, with the use of an inpainting algorithm \citep{Mathur10,Pires15} to minimize the effect of up to 2\,day-long gaps, during regular Earth downlinks and angular momentum dumps. A color-magnitude diagram for all the stars in the sample is shown in Fig.~\ref{fig:cmd}, which emphasizes the average difference in mass among the three open clusters (see also \citealt{Corsaro12} for more details about the general properties of the population of RGs in these open clusters). We notice that two stars marked as RGB, namely KIC~2437589 in NGC\,6791 and KIC~5112361 in NGC\,6819, are placed in the region of the color-magnitude diagram where the corresponding RC stars of the same clusters are located. Despite their peculiar location in the diagram, both stars have a RGB evolutionary stage unambiguously determined by their oscillations \citep{Corsaro12,Corsaro17}, with KIC~2437589 a possible evolved blue straggler \citep{Brogaard12,Corsaro12} and KIC~5112361 a spectroscopic single lined binary \citep{Milliman14}.

\subsection{Effective temperatures}
\label{sec:temperature} 
For obtaining an accurate set of stellar effective temperatures for the entire sample of stars in this study we start from the revised KIC temperatures from the Sloan Digital Sky Survey \citep[SDSS,][]{Gunn06SDSS} \emph{griz} filters \citep{Pinsonneault12SDSS}, which are available for all 60 targets. In addition, 36 stars (12 in NGC\,6791, 20 in NGC\,6819, and 4 in NGC\,6811) have new temperatures determined from spectroscopy with ASPCAP \citep[APOGEE Stellar Parameters and Chemical Abundances Pipeline,][]{Zasowski13APOGEE,Nidever15APOGEE,Holtzman15APOGEE,GarciaPerez16APOGEE,Majewski17APOGEE}, using the Data Release 13 (DR13, \citealt{Albareti16SDSS}) of SDSS IV \citep{Blanton17SDSS}, which includes the post-release metallicity correction (see Holtzman et al. in prep.). We therefore use ASPCAP temperatures, available from the latest release of the APOKASC catalog \citep{Pinsonneault14APOKASC,Tayar17}, to apply a zero point shift to the temperatures from SDSS and correct them for the different cluster extinctions, which were based on the KIC map \citep{Brown11KIC} in the work by \cite{Pinsonneault12SDSS}. In this way we put the temperatures from SDSS on the same scale as ASPCAP and we adopt the typical ASPCAP total temperature uncertainty (including both systematic and random effect) of $\sim69\,$K as a reference (see Holtzman et al. in prep. and \citealt{Tayar17} for more discussion). 

\begin{figure}
   \centering
   \includegraphics[width=9.2cm]{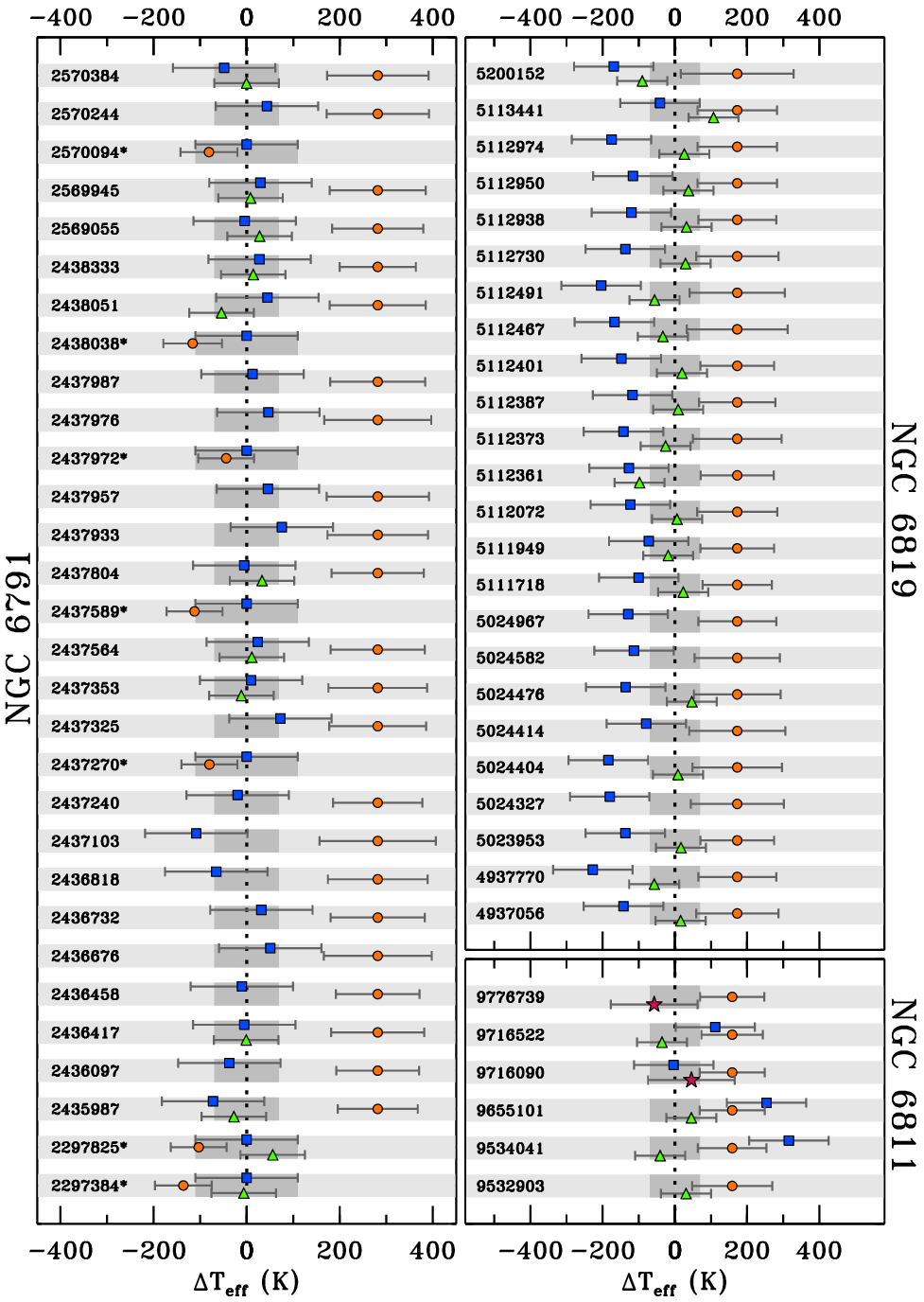}
      \caption{Different sources for $T_\mathrm{eff}$ as compared to the SDSS-based temperature scale prepared in this work and taken as a reference ($\Delta T_\mathrm{eff} = 0$\,K is marked with a dotted line), for all the 60 cluster RGs, labeled with their KIC ID. Orange circles refer to the original temperatures from the SDSS photometry \citep{Pinsonneault12SDSS}, blue squares to $(V-K)$ color temperatures \citep{Hekker11OC}, green triangles to ASPCAP temperatures from APOKASC \citep{Pinsonneault14APOKASC}, and purple stars to spectroscopic temperatures from \cite{Molenda14NGC6811}. The 1-$\sigma$ uncertainties on each value are also overlaid. The dark gray shading around $\Delta T_\mathrm{eff} = 0$ delimits the 1-$\sigma$ uncertainty adopted on the reference SDSS-based temperature scale. The seven stars of NGC\,6791 that show cooler SDSS temperatures (see Sect.~\ref{sec:temperature}) are marked by an asterisk.}
    \label{fig:teff_comparison}
\end{figure}

From a detailed comparison of individual temperature values, we noticed that several stars in NGC\,6791 (specifically KIC\,2297384, KIC\,2297825 on the RC, and KIC\,2437270, KIC\,2437589, KIC\,2437972, KIC\,2438038, KIC\,2570094 on the RGB) have SDSS temperatures that are systematically cooler (by about 374\,K) than the average SDSS temperatures of the other red giants in the same cluster. This discrepancy is clearly visible by looking at the corresponding temperature differences shown with orange circles in Fig.~\ref{fig:teff_comparison}, where the seven stars that we mentioned are marked by an asterisk. This $\sim374$\,K offset is caused by an adopted reddening for the seven stars that is smaller than that of the other cluster stars by about $0.2$. For these seven stars we therefore decided to use $(V-K)$ color temperatures (and corresponding uncertainties of 110\,K, \citealt{Hekker11OC}), which are available for all targets in NGC\,6791. This choice is motivated by the fact that the $(V-K)$ color temperatures for the stars in NGC\,6791 are in agreement (well within 1-$\sigma$) with the ASPCAP temperatures from APOKASC and with those from our new temperature scale (see the comparison in Fig.~\ref{fig:teff_comparison}). Finally, to avoid biasing our extinction correction applied to the SDSS temperature scale using the ASPCAP one, we remove KIC\,2297384 and KIC\,2297825 (the only two stars out of the seven with cool SDSS temperature that have also an ASPCAP temperature, see Fig.~\ref{fig:teff_comparison}) from the computation of the zero point shift. The final temperature shifts that we obtain are $\langle T_\mathrm{eff, SDSS} - T_\mathrm{eff,ASPCAP} \rangle = 282$\,K for NGC\,6791, 173\,K for NGC 6819, and 156\,K for NGC 6811, showing that temperatures from SDSS photometry are systematically hotter than the ASPCAP ones (see Fig.~\ref{fig:teff_comparison}). For simplicity, from here onwards the so-called \emph{SDSS-based} temperature scale will refer to the temperatures from SDSS photometry corrected to the ASPCAP temperature scale as explained in this section, and supplemented with $(V-K)$ color temperatures adopted for the seven targets in NGC\,6791 that show SDSS temperatures 400\,K cooler than the other stars in the same cluster. We therefore adhere to the SDSS-based temperature scale to compute corrected mass estimates as discussed in Sect.~\ref{sec:mass}. A complete list of the adopted temperatures for each star in the sample can be found in Tables~\ref{tab:stellar_parameters_NGC6791}, \ref{tab:stellar_parameters_NGC6819}, \ref{tab:stellar_parameters_NGC6811}. 

For completeness, we also note that $(V-K)$ color temperatures are available for all of the stars in NGC\,6819 and in NGC\,6811, except KIC~9776739 and KIC~9716090, which were studied by \cite{Molenda14NGC6811} and have temperatures from spectroscopic data acquired from the Nordic Optical Telescope. As visible from Fig.~\ref{fig:teff_comparison}, for the stars in NGC\,6791 we find a good agreement between $(V-K)$ color temperatures and ASPCAP temperatures, while this agreement partially weakens for the stars in NGC\,6819 where $(V-K)$ color temperatures are systematically cooler, and in NGC\,6811 where instead they are hotter, although compatibility between the difference sources is still ensured within 1-$\sigma$ in most cases. Lastly, the spectroscopic measurements from \cite{Molenda14NGC6811} also agree (well within 1-$\sigma$) with our SDSS-based temperatures for the same stars (Fig.~\ref{fig:teff_comparison}).

\subsection{Metallicity}
\label{sec:metallicity}
We consider new cluster mean metallicities computed from ASPCAP corrected metallicities for the 36 targets for which they are available (see also Sect.~\ref{sec:temperature}, and \citealt{Tayar17}). The metallicity values that we obtain for each cluster are listed in Table~\ref{tab:cluster_feh}, with an uncertainty showing the standard deviation on the mean of the sample. Our estimates show that NGC\,6791 has about twice the solar metallicity, NGC\,6819 a close-to-solar metallicity, while NGC\,6811 has a slightly sub-solar one (about 20\,\% less than that of the Sun). These cluster mean metallicities are in good agreement with previous estimates found in the literature for \ngca, [Fe/H]\,$= 0.29 \pm 0.08$ \citep{Brogaard11}, \ngcb, [Fe/H]\,$= 0.09 \pm 0.03$ \citep{Bragaglia01OC} and \ngcc, [Fe/H]\,$= -0.040 \pm 0.002$ \citep{Molenda14NGC6811}. We provide individual metallicity values from ASPCAP in Tables~\ref{tab:stellar_parameters_NGC6791}, \ref{tab:stellar_parameters_NGC6819}, \ref{tab:stellar_parameters_NGC6811}, but we will consider only the cluster mean metallicities in the analysis presented in Sect.~\ref{sec:inference}, to exploit the common origin that characterizes the stars in clusters. The metallicity range covered by the stars in the three clusters, about $\sim0.4$\,dex, while not large, is sufficient to identify the effects of metallicity on stellar granulation with high reliability and statistical evidence thanks to the homogeneity of the stellar properties shared by the members of each cluster (see also Sect.~\ref{sec:discussion} for more discussion).

\begin{table}
\centering
 \caption{Cluster mean metallicities and corresponding uncertainties as computed from the available ASPCAP metallicities for 36 stars of the sample investigated in this work (see also Tables~\ref{tab:stellar_parameters_NGC6791}, \ref{tab:stellar_parameters_NGC6819}, \ref{tab:stellar_parameters_NGC6811}, for a list of all the values).}
\begin{tabular}{lr}
\hline\hline
\\[-8pt]
Open cluster & \multicolumn{1}{c}{$\langle \mbox{[Fe/H]} \rangle_\mathrm{ASPCAP}$} \\[1pt]
  & \multicolumn{1}{c}{(dex)} \\ [1pt]
 \hline
 \\[-8pt]       
   NGC\,6791 & $0.32 \pm 0.02$ \\ [1pt]
   NGC\,6819 & $0.04 \pm 0.03$ \\ [1pt]
   NGC\,6811 & $-0.09 \pm 0.03$ \\ [1pt]
  \hline
 \end{tabular}
\label{tab:cluster_feh}
\end{table}

\subsection{Stellar mass}
\label{sec:mass}
\begin{figure}
   \centering
   \includegraphics[width=9.0cm]{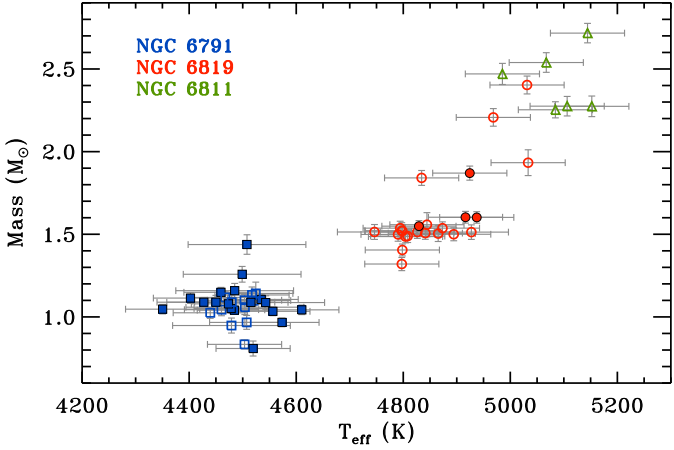}
      \caption{Corrected stellar masses from Sect.~\ref{sec:mass}, for the 60 cluster RGs, as a function of the temperatures from the SDSS-based temperature scale. The same symbol coding as in Fig.~\ref{fig:cmd} is adopted. The 1-$\sigma$ uncertainties in mass and $T_\mathrm{eff}$ are also overlaid.}
    \label{fig:teff_mass}
\end{figure}

We estimate the stellar masses and their uncertainties by basing our analysis on the asteroseismic scaling relations (e.g., \citealt{Miglio12massloss}). For RGs especially, it is recognized that the asteroseismic scaling relations have the tendency to overestimate masses because scaling relations are approximate in nature \citep[e.g.,][]{Brown91numax,Belkacem11}. To compensate for the overestimation, many authors have proposed different corrections, some empirically-based \citep{Mosser13}, others resulting from calibrations using stellar evolution models \citep{White11, Miglio12massloss, Sharma16, Guggenberger16}. In addition, \citet{Gaulme16} compared masses and radii, obtained independently from both asteroseismic relations and dynamical modeling (eclipse photometry combined with radial velocities), of a sample of RGs in eclipsing binary systems. As a result, asteroseismic masses appear to be about 15\,\% larger than dynamical masses (see \citealt{Gaulme16} Fig. 9). However, since the sample studied by \citet{Gaulme16} is rather small, we cannot infer how the mass correction depends on stellar parameters to provide a general correction law. Instead we follow \cite{Sharma16} to compute a correction factor to the scaling relation of the large frequency separation $\Delta\nu$ \citep{Ulrich86}. This correction is based on a large grid of stellar evolution models. We therefore adopt a modified version of the standard scaling relation for mass, which reads as
\begin{equation}
\frac{M}{M_{\odot}} = \left( \frac{\numax}{\nu_\mathrm{max,\odot}} \right)^3 \left( \frac{\Dnu}{\gamma \Dnu_{\odot}} \right)^{-4} \left( \frac{T_\mathrm{eff}}{T_\mathrm{eff,\odot}} \right)^{1.5} \, ,
\end{equation}
where $\gamma$ is a correction factor for $\Dnu$, and is computed by taking into account the values of temperature, $\numax$, and $\Dnu$ for each star, and the cluster mean metallicities from Table~\ref{tab:cluster_feh}. Uncorrected mass estimates (from pure scaling) can easily be recovered with $M_\mathrm{uncorr} = M \gamma^{-4}$. The value of $\Dnu$ for each star is computed from the frequencies of the three radial modes that are closest to $\numax$. The frequencies are obtained from the peak bagging analysis performed by Corsaro et al. (in prep.), which consists  in the fitting and identification of individual oscillation modes to extract their frequencies, amplitudes, and lifetimes. The peak bagging analysis for the cluster RGs of our sample is done following the same recipe presented by \cite{Corsaro15cat} (see also \citealt{Corsaro17}), and by adopting the background parameters estimated in this work. We refer the reader to \cite{Corsaro15cat} for a detailed description of the peak-bagging analysis process using the Bayesian inference code \diamonds\,\,\citep{Corsaro14}. The resulting stellar masses using the SDSS-based temperature scale (Sect.~\ref{sec:temperature}) and the cluster mean metallicities (Table~\ref{tab:cluster_feh}) are listed in Tables~\ref{tab:stellar_parameters_NGC6791}, \ref{tab:stellar_parameters_NGC6819}, \ref{tab:stellar_parameters_NGC6811}, while the solar reference values $\nu_\mathrm{max,\odot}$, $\Dnu_{\odot}$, $T_\mathrm{eff,\odot}$, are presented in Sect.~\ref{sec:virgo}. By defining the mass difference between corrected and uncorrected estimates, $\Delta M = M - M_\mathrm{uncorr}$, we note that the average mass correction $\langle \Delta M / M_\mathrm{uncorr} \rangle$ is $-6.5$\,\% for NGC\,6791, $0.5$\,\% for NGC\,6819, and $4.0$\,\% for NGC\,6811. In Fig.~\ref{fig:teff_mass} we show our estimates of stellar mass as a function of $T_\mathrm{eff}$ from the SDSS-based temperature scale, where a clear correlation between these two parameters is found, especially at temperatures higher than 4600\,K.

\section{Analysis of the background signal}
\subsection{Background fitting model}
\label{sec:bkg}
The starting point of the analysis presented in this work is the measurement of the background properties observed in the stellar power spectral densities (PSDs) obtained from the \kepler light curves. We determine the parameters of the background signal (including granulation), as detailed in \cite{Corsaro15cat}, using the Bayesian inference code \diamonds\,\,\citep{Corsaro14}. We adopt the background model presented by K14 \citep[see also][]{Kallinger16Science}, which can be expressed as
\begin{equation}
P_\mathrm{bkg} \left(\nu \right) = N\left( \nu \right) + R \left( \nu \right) \left[ B\left( \nu \right) + G \left( \nu \right) \right] \, ,
\label{eq:overall_bkg}
\end{equation}
where we assume the noise component
\begin{equation}
N(\nu) = W + \frac{2 \pi a_\mathrm{n}^2/b_\mathrm{n}}{1 + \left(\nu/b_\mathrm{n}\right)^2} \, ,
\label{eq:noise}
\end{equation}
to be the combination of a flat noise level $W$, mainly dominant at high frequency ($\nu \sim 200$\,\muhz), and a colored noise that can become significant at low frequency ($\nu \leq 20\,\mu$Hz), with $a_\mathrm{n}$ the amplitude, $b_\mathrm{n}$ its characteristic frequency, and $2\pi$ a normalization constant (see also K14). The three super-Lorentzian components
\begin{equation}
B\left(\nu\right) = \sum^3_{i=1} \frac{\zeta a^2_i / b_i}{1 + \left( \nu / b_i \right)^4} \, ,
\label{eq:bkg}
\end{equation}
describe in decreasing frequency order, the granulation at frequencies close to $\numax$, the meso-granulation for frequencies close to $\numax/3$, and a low-frequency component that incorporates heterogeneous signal coming from possible super-granulation and low-frequency variations in the time-series, whose analysis is beyond the scope of this work. Here $a_i$ is the rms intensity fluctuation (or amplitude), $b_i$ the characteristic frequency, and $\zeta = 2\sqrt{2}/\pi$ the normalization constant for a super-Lorentzian profile with its exponent set to four \citep[see][K14 for more details]{Karoff13}. The power excess containing the stellar oscillations is modeled using a Gaussian envelope defined as
\begin{equation}
G\left(\nu\right) = H_\mathrm{osc} \exp \left[ - \frac{ \left( \nu - \nu_\mathrm{max} \right)^2}{2 \sigma_\mathrm{env}^2} \right] \, ,
\label{eq:env}
\end{equation}
with $H_\mathrm{osc}$ the height of the oscillation bump and $\sigma_\mathrm{env}$ the standard deviation. Finally, all the components of the background signal, except those purely related to noise, $N(\nu)$, are modulated by the response function that corrects for the finite integration time of the long cadence \kepler observations, expressed as
\begin{equation}
R\left( \nu \right) = \mbox{sinc}^2 \left( \frac{\pi \nu}{2 \nu_\mathrm{Nyq}} \right) \, ,
\label{eq:resp}
\end{equation}
with $\nu_\mathrm{Nyq} = 283.212\,\mu$Hz the associated Nyquist frequency. 

In Eq.~(\ref{eq:bkg}), the meso-granulation component is associated with the parameters $(a_2, b_2)$, while the granulation component corresponds to the parameters $(a_3, b_3)$, with $a_2 > a_3$ and $b_2 < b_3$. The granulation component is the one that can be modeled through existing 3D hydrodynamical simulations of stellar atmospheres \citep[e.g., see][]{Trampedach98,Ludwig16}. However, in this work we will focus our analysis on the meso-granulation component, and we will refer to it from now on using the symbols $a_\mathrm{meso} \equiv a_2$ and $b_\mathrm{meso} \equiv b_2$. We will occasionally refer to the granulation component using the symbols $a_\mathrm{gran} \equiv a_3$ and $b_\mathrm{gran} \equiv b_3$. We have decided to select and analyze the meso-granulation component for the following reasons: (i) it is dominant over the granulation component both in height (PSD units) and in amplitude \citep[e.g., see K14,][]{Corsaro15cat}, so it is statistically more significant; (ii) it is well detached from the oscillation bump (with $\nu_\mathrm{max} \approx 3 b_\mathrm{meso}$), hence less affected by biases and correlations associated with stellar oscillations than the granulation component, for which $\nu_\mathrm{max} \approx b_\mathrm{gran}$ \citep[e.g., see][]{Corsaro14}; (iii) its characteristic parameters can be better determined than those of the granulation due to a higher signal-to-noise ratio, especially at high $\numax$ (e.g., $> 100\,\mu$Hz); (iv) it scales to surface gravity and temperature of the star similarly to the granulation component, because the meso-granulation represents a reorganization of the granulation at larger scales, so it originates from the same envelope convective motions. Hence, $a_\mathrm{meso}$ and $b_\mathrm{meso}$ can be used as accurate proxies for $a_\mathrm{gran}$ and $b_\mathrm{gran}$ (see K14). On average we find that $a_\mathrm{meso}/a_\mathrm{gran} = 1.31 \pm 0.18$ and that $b_\mathrm{meso}/b_\mathrm{gran} = 0.32 \pm 0.04$, throughout the $\numax$ range spanned by our stellar sample.

\begin{figure}
   \centering
   \includegraphics[width=9.0cm]{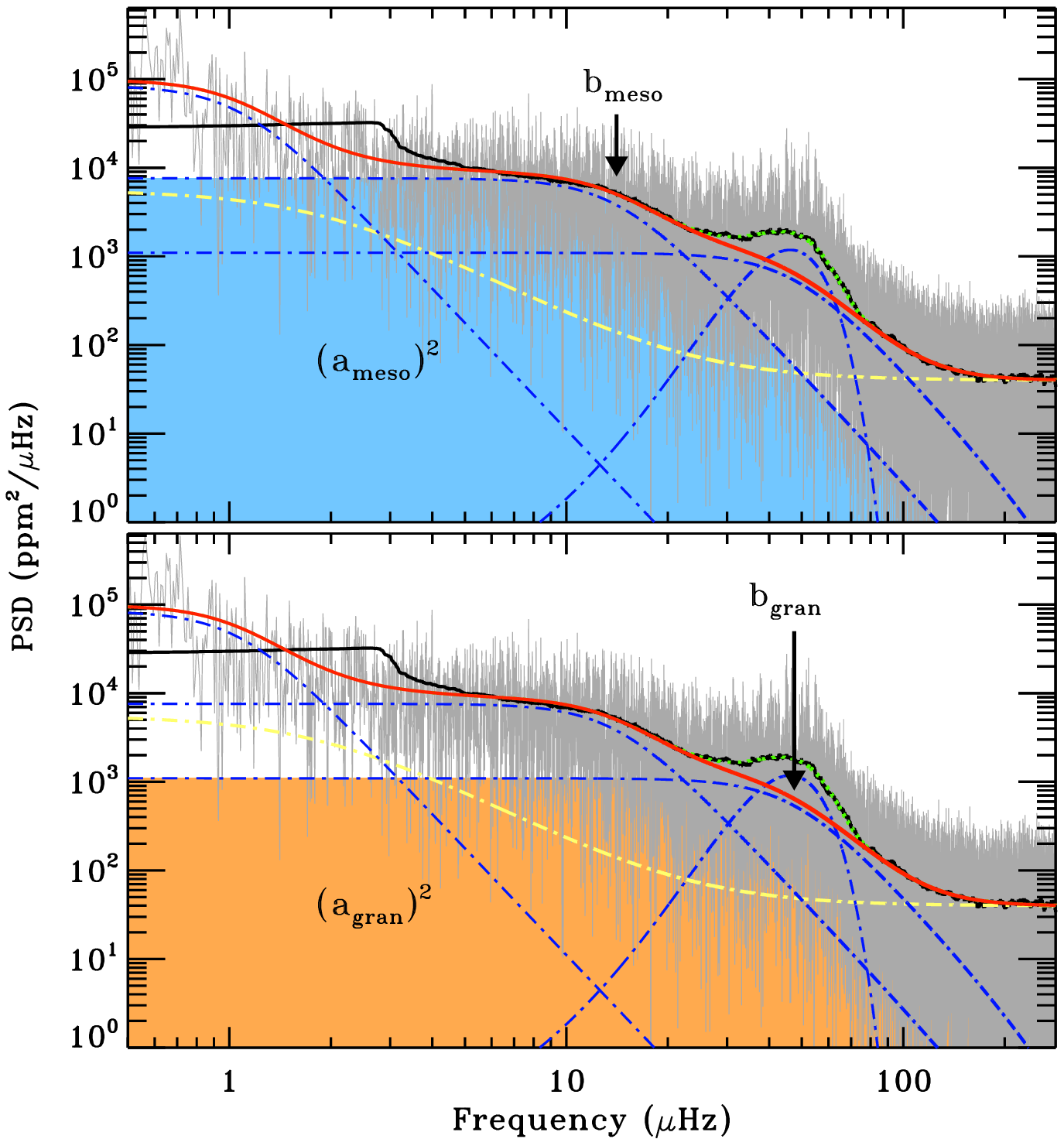}
      \caption{Resulting background fit done with \diamonds\,\,(red curve) for the star KIC~4937056 in the cluster NGC\,6819 overlaid on the original PSD of the star (in gray) and its smoothed version (black curve) using a boxcar with a width set to $\Dnu/5$, where $\Dnu$ is computed using the $\Dnu-\numax$ relation for RGs calibrated by \cite{Huber11}. The different components that constitute $B(\nu)$ (Eq.~\ref{eq:bkg}) and $G(\nu)$ (Eq.~\ref{eq:env}) are indicated with dot-dashed blue lines, while the noise term $N(\nu)$ (Eq.~\ref{eq:noise}) is shown with a dot-dashed yellow curve. The dotted green curve shows the overall fit of the background when the Gaussian envelope is included. Median values of the free parameters for each background component are used, as listed in Table~\ref{tab:bkg_parameters_NGC6819}. \textit{Top panel}: the meso-granulation component $(a_\mathrm{meso}, b_\mathrm{meso})$ appears as a kink at $\sim10$\,\muhz\,\,(arrow) and its amplitude squared ($a_\mathrm{meso}^2$) is represented by the area of the shaded blue region. \textit{Bottom panel}: same as the top panel but showing the parameters of the granulation component $(a_\mathrm{gran}$, $b_\mathrm{gran})$.}
    \label{fig:bkg_case}
\end{figure}

\begin{figure}
   \centering
   \includegraphics[width=9.0cm]{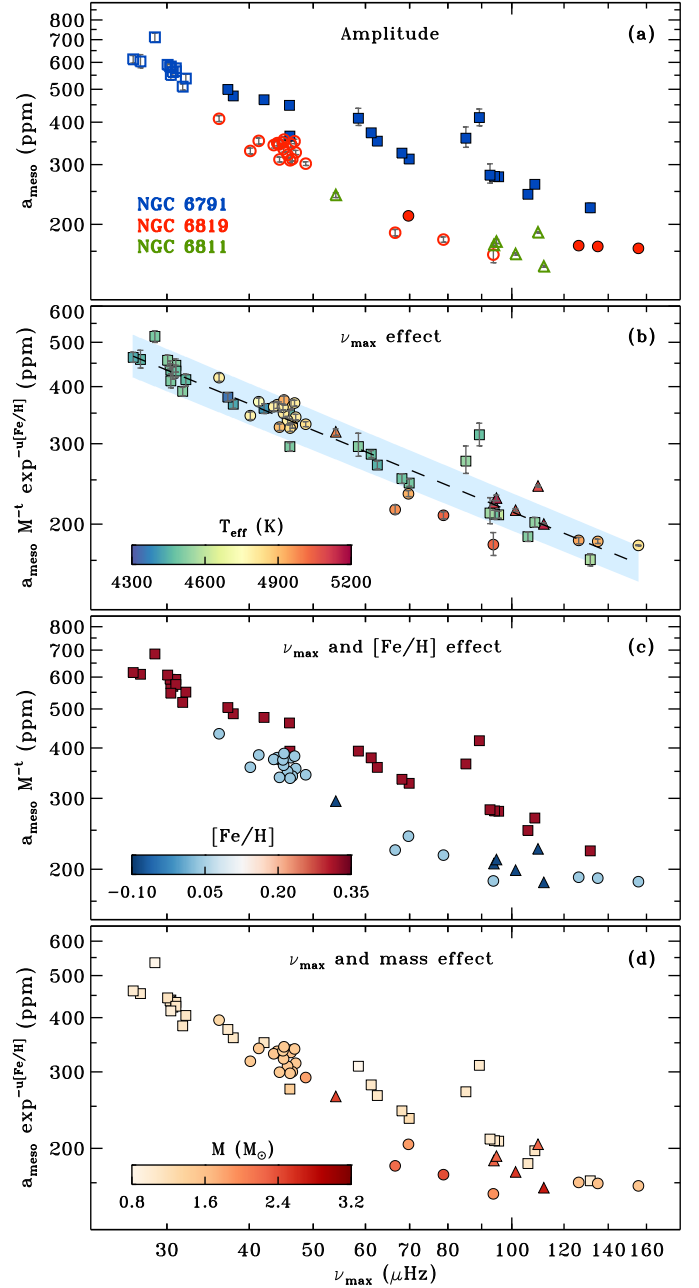}
      \caption{Amplitude of the meso-granulation component as a function of $\nu_\mathrm{max}$ for the 60 cluster RGs. Color-coding in panel (a) shows the cluster membership, with open symbols for RC and filled ones for RGB stars, and the symbol types indicating the cluster membership as in Fig.~\ref{fig:teff_mass}. Panel (b) shows the amplitudes after removing the effect of both mass and metallicity using the best scaling relation identified in Sect.~\ref{sec:results} (model $\mathcal{M}_{a,2}$), with the resulting fit marked by a dashed line and corresponding 1-$\sigma$ credible region in shading, and with $T_\mathrm{eff}$ from the SDSS-based temperature scale color-coded for each star. Panels (c) and (d) show the amplitudes after removing the effect of only mass and only metallicity, respectively, where cluster mean metallicities from ASPCAP and corrected masses of the stars from Sect.~\ref{sec:mass} are color-coded. Bayesian credible intervals of 68.3\% on meso-granulation amplitudes are shown in panel (a), and are rescaled in panel (b).}
    \label{fig:amp_numax}
\end{figure}

\begin{figure}
   \centering
   \includegraphics[width=9.0cm]{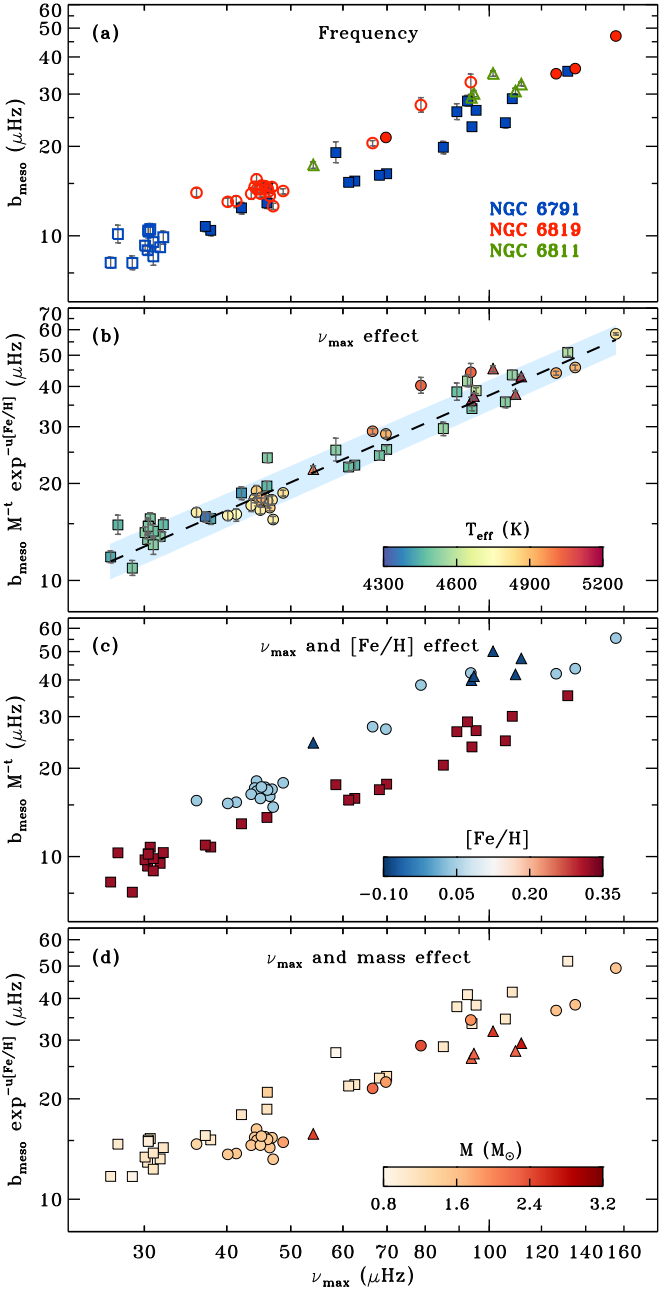}
      \caption{Same as Fig.~\ref{fig:amp_numax} but for the meso-granulation characteristic frequency.}
    \label{fig:freq_numax}
\end{figure}

Figure~\ref{fig:bkg_case} shows an example of the resulting fit with \diamonds\,\,using the model given by Eq.~(\ref{eq:overall_bkg}) for the cluster RG KIC~4937056. The meso-granulation parameters and $\nu_\mathrm{max}$ from the fit to all the stars are presented in Figs.~\ref{fig:amp_numax}a and \ref{fig:freq_numax}a, and listed in Tables~\ref{tab:bkg_parameters_NGC6791},  \ref{tab:bkg_parameters_NGC6819} and \ref{tab:bkg_parameters_NGC6811}. We use uniform priors for all free parameters of the background model. The uniform prior boundaries are obtained by performing preliminary fits with the automated pipeline A2Z \citep{Mathur10}, and using $\numax$ values from \cite{Corsaro12} and from \cite{Molenda14NGC6811} as additional inputs. The configuring parameters of \diamonds\,\,that are adopted for the fitting of the background signal are provided in Appendix~\ref{sec:bkg_results}. 

\begin{figure}
   \centering
   \includegraphics[width=9.0cm]{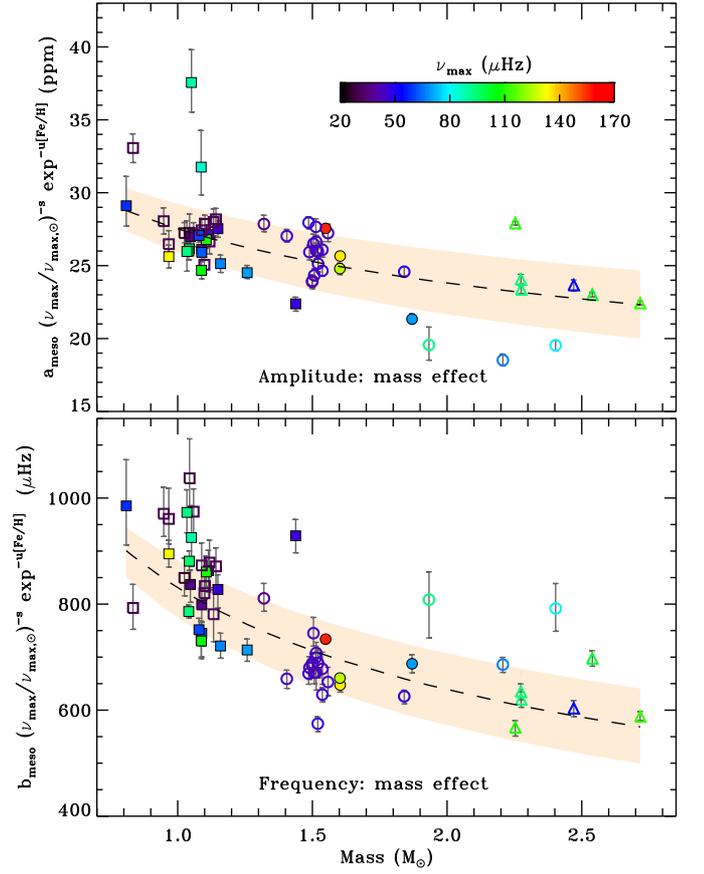}
      \caption{Amplitudes (top panel) and frequencies (bottom panel) of the meso-granulation component for the 60 cluster RGs, as a function of the corrected stellar masses from Sect.~\ref{sec:mass}. Dependencies upon $\numax$ and [Fe/H] were removed by means of the best scaling relations identified in Sect.~\ref{sec:model_comparison} (Eq.~\ref{eq:scal_amp_meso} for the top panel and Eq.~\ref{eq:scal_freq_meso} for the bottom panel). The dashed lines mark the fits from the best scaling relations, while the shaded regions delimit the 1-$\sigma$ credible regions from the estimated parameters (Table~\ref{tab:param_gran}). Open symbols denote RC stars, while filled symbols are RGB stars, with cluster membership indicated by the same symbol types as in Fig.~\ref{fig:teff_mass}. The values of $\numax$ are color-coded for each star. Rescaled Bayesian credible intervals of 68.3\,\% for each meso-granulation parameter are overlaid on both panels.}
    \label{fig:meso_mass}
\end{figure}

\begin{figure}
   \centering
   \includegraphics[width=9.0cm]{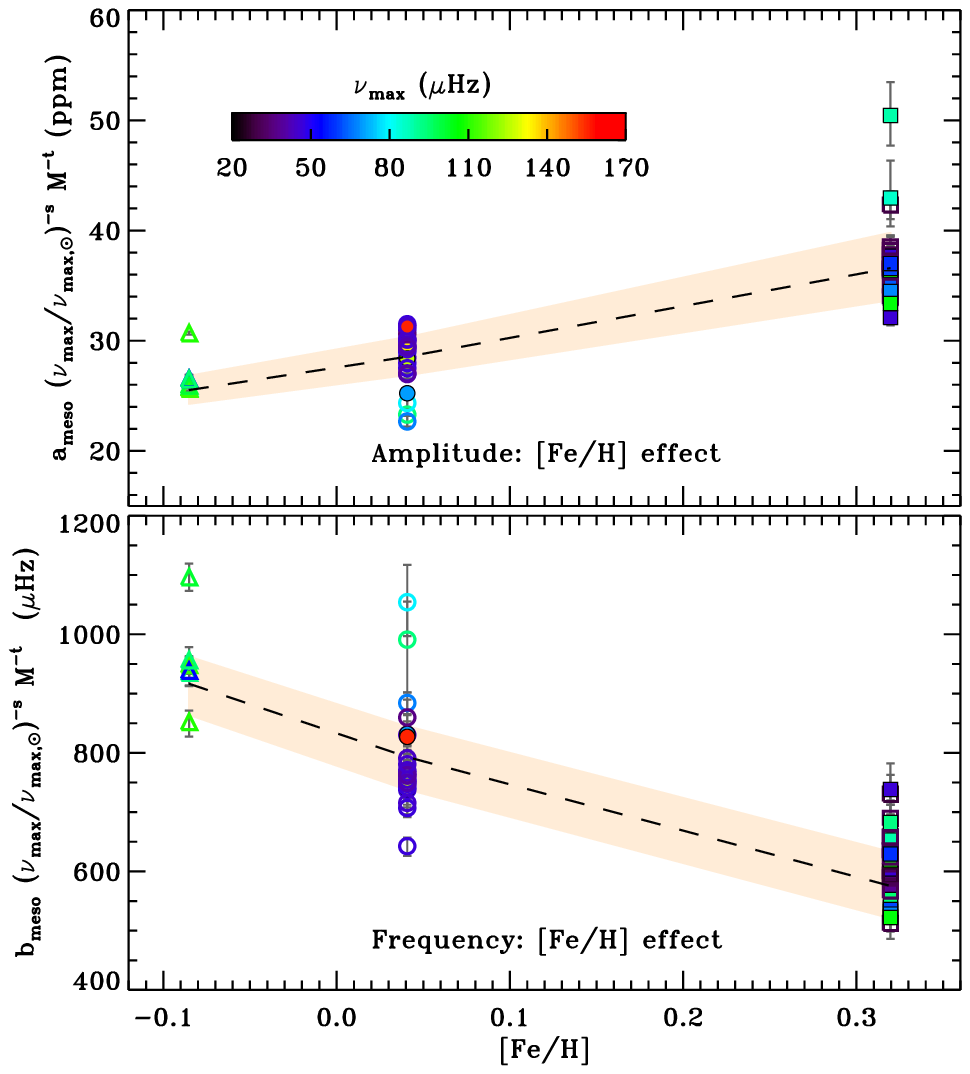}
      \caption{Same as Fig.~\ref{fig:meso_mass} but showing the meso-granulation parameters with the $\numax$ and stellar mass dependencies removed, as a function of the cluster mean metallicities computed from ASPCAP (Sect.~\ref{sec:metallicity}).}
    \label{fig:meso_feh}
\end{figure}

\subsection{Solar reference values from VIRGO}
\label{sec:virgo}
For a proper assessment of any metallicity dependence on granulation as presented in Sect.~\ref{sec:scal_rel}, we rely on our derivation of solar reference values. These reference values need to be as accurate and consistent as possible with the photometric data used for the cluster RGs presented in Sect.~\ref{sec:photometry} and the analysis described in Sect.~\ref{sec:bkg}. For this purpose, we apply the background fitting approach with \diamonds\,\,(see Sect.~\ref{sec:bkg}) to the PSD of the Sun, using the background model defined by Eq.~(\ref{eq:overall_bkg}). We consider the combined light curves from VIRGO green and red channels $(g+r)$ to mimic the broad \kepler bandpass \citep{Basri10VIRGOasKepler}. We use an observing length coinciding with that of the \kepler light curves used in this work (see Sect.~\ref{sec:photometry}), thus obtaining the same frequency resolution in the resulting PSD. We consider two different combined light curves, the first one centered around the maximum of solar activity and the second one centered around the minimum. This is done to average out the effect of the solar activity cycle on the observed properties of the Sun. The two PSDs are then computed in the same way as for \kepler stars, and by re-binning to a sampling rate of 60\,s, close to that of the \kepler short cadence observation. We obtain two sets of solar parameters, one corresponding to maximum solar activity and the other to minimum solar activity, which we average to obtain final estimates that are not biased by the activity cycle of the Sun. The final solar reference values are $a_\mathrm{meso,\odot} = 56.0 \pm 0.2\,$ppm (parts-per-milion) and $b_\mathrm{meso,\odot} = 752 \pm 3\,\mu$Hz for the meso-granulation component, and $\nu_\mathrm{max,\odot} = 3147 \pm 2\,\mu$Hz for the power excess due to solar oscillations. We also include the large frequency separation as the average from the two datasets, $\Delta\nu_{\odot} = 135.04 \pm 0.02\,\mu$Hz, whose calculation is explained in detail by Corsaro et al. (in prep.) and follows from a similar approach as that applied to the cluster RGs. Finally, the reference value for the solar effective temperature is the typical $T_\mathrm{eff} = 5777$\,K \citep[e.g.,][]{Corsaro13}. Our estimates of $b_\mathrm{meso,\odot}$ and $\nu_\mathrm{max,\odot}$ agree within $1$-$\sigma$ and $2$-$\sigma$, respectively, with those obtained by K14. Our value of $a_\mathrm{meso,\odot}$ is instead about 1.5 times larger than that of K14. We attribute this difference in $a_\mathrm{meso,\odot}$ to the different preparation of the solar dataset, which in the case of K14 was accounting for 1-year length observation of the VIRGO green channel only (centered at 550\,nm), and thus applying a simple linear transformation to obtain the reference amplitude at the central wavelength of the \kepler bandpass (664\,nm).

\section{Scaling relations for granulation activity}
\label{sec:scal_rel}
So far, empirical models related to the efficiency of the granulation signal, represented by the granulation amplitude $a_\mathrm{gran}$ and its characteristic frequency $b_\mathrm{gran}$ --- or equivalently its time scale $\tau_\mathrm{gran} = (2\pi b_\mathrm{gran})^{-1}$ --- have been investigated using large samples of stars with evolutionary stages ranging from the main sequence to the late RG phase \citep[][K14]{Mathur11granulation,Kallinger16Science}. In the following we present the relevant scaling relations connecting $a_\mathrm{meso}$ and $b_\mathrm{meso}$ to $\numax$, which in turn depends on stellar surface gravity and temperature, and mass of the stars, and for the first time we include the additional dependence on stellar metallicity.

\subsection{Meso-granulation amplitude $a_\mathrm{meso}$}
\label{sec:scal_amp}
As shown originally by \cite{Mathur11granulation}, and later on by K14 and by \cite{Kallinger16Science}, the stellar granulation signal is strongly correlated with the atmospheric parameters of effective temperature and surface gravity, therefore $\nu_\mathrm{max}$ \citep{Brown91numax}. For constant surface gravity, one can also test the effect of a varying stellar mass. In this work we consider a more general scaling relation of the form
\begin{equation}
\left( \frac{a_\mathrm{meso}}{a_\mathrm{meso,\odot}} \right) = \beta \left( \frac{\nu_\mathrm{max}}{\nu_\mathrm{max,\odot}} \right)^s \left(\frac{M}{M_{\odot}} \right)^t e^{u {\left[ \rm{Fe/H} \right]}} \, ,
\label{eq:scal_amp_meso}
\end{equation}
where $M$ is the mass of the star, [Fe/H] the metallicity, $s$, $t$, and $u$ are three exponents that need to be estimated, and $\beta$ is a scaling parameter for the solar reference values, typically set to $\beta = 1$ \citep[see also][for more discussion about the implications of this scaling factor]{Corsaro13}. We note that the solar values are only reference values used for the scalings, and a different choice of these values will not change either the quality of the fits or the Bayesian evidence associated with each model (see Sect.~\ref{sec:model_comparison}), but it would instead impact on the term $\beta$ that calibrates the scaling relation. As discussed in Sect.~\ref{sec:virgo}, we use our own solar reference values for consistency. We parametrize the metallicity with an exponential function because [Fe/H] is already expressed in a logarithmic form and is compared to the solar value. For the study presented here, we consider two cases deriving from the generalized scaling relation of Eq.~(\ref{eq:scal_amp_meso}), one for $u = 0$, which only accounts for $\numax$ and mass dependencies, and one for $u \ne 0$, which also includes the metallicity effect on the amplitudes. We decide not to investigate the scaling relations incorporating only the dependency on $\numax$ (obtained for $t = 0$ and $u = 0$) and only the dependency on $\numax$ and [Fe/H] (obtained for $t = 0$), because the effect of a varying stellar mass in the amplitudes was already found to be significant from previous analyses (see K14).

In order to linearize the scaling relations and to be able to perform a thorough statistical analysis following the approach shown by \cite{Corsaro13} and \cite{Bonanno14}, we apply the natural logarithm to  Eq.~(\ref{eq:scal_amp_meso}), yielding 
\begin{equation}
\ln \left( \frac{{a}_\mathrm{meso}}{a_\mathrm{meso,\odot}} \right) = \ln \beta + s \ln \left( \frac{\nu_\mathrm{max}}{\nu_\mathrm{max,\odot}} \right) + t  \ln \left(\frac{M}{M_{\odot}} \right) + u \left[ \rm{Fe/H} \right] \, .
\end{equation}
From here onwards, we identify the linearized scaling relations (or models) for the meso-granulation amplitude with the symbols $\mathcal{M}_{a,1}$ for $u = 0$, and $\mathcal{M}_{a,2}$ for $u \ne 0$ (see Sect.~\ref{sec:inference} for more details). We also compute the analytical expressions for the uncertainties associated to the predicted meso-granulation amplitudes. According to a standard Gaussian error propagation (see also \citealt{Corsaro13}), we obtain the total relative uncertainties
\begin{equation}
\widetilde{\sigma}^2_{a} (s,t,u) = \widetilde{\sigma}_{a_\mathrm{meso}}^2 + s^2 \widetilde{\sigma}_{\nu_\mathrm{max}}^2 + t^2 \widetilde{\sigma}_M^2 + u^2 \widetilde{\sigma}_{\rm{[Fe/H]}}^2 \, ,
\end{equation}
which clearly depend upon the free parameters of the corresponding scaling relations, except for the offset term $\ln \beta$ that is not directly depending on any of the observables in this formulation. The relative uncertainties are $\widetilde{\sigma}_{a_\mathrm{meso}} \equiv \sigma_{a_\mathrm{meso}}/a_\mathrm{meso}$, $\widetilde{\sigma}_{\nu_\mathrm{max}} \equiv \sigma_{\nu_\mathrm{max}} / \nu_\mathrm{max}$, $\widetilde{\sigma}_M \equiv \sigma_M / M$, while $\widetilde{\sigma}_{\rm{[Fe/H]}}$ is the formal uncertainty on metallicity (which is already in relative units). Clearly, the total relative uncertainty for amplitude predictions from model $\mathcal{M}_{a,1}$ is obtained by imposing $u = 0$, while that from model $\mathcal{M}_{a,2}$ is the general form with $u \neq 0$. These parameter-dependent uncertainties, and the linearized models, will be used for the Bayesian inference described in Sect.~\ref{sec:inference}.

\subsection{Meso-granulation characteristic frequency $b_\mathrm{meso}$}
\label{sec:scal_freq}
The frequency scale of the granulation signal is known to follow a tight scaling with the stellar surface gravity, like the amplitude. In particular \cite{Mathur11granulation} showed that the time scale of the granulation signal, $\tau_\mathrm{gran}$, scales with $\nu_\mathrm{max}$. This result was later on confirmed by K14. For our analysis we adopt a generalized scaling relation for the characteristic frequency of the meso-granulation signal, of the form
\begin{equation}
\left( \frac{b_\mathrm{meso}}{b_\mathrm{meso,\odot}} \right) = \beta \left( \frac{\nu_\mathrm{max}}{\nu_\mathrm{max,\odot}} \right)^s \left( \frac{M}{M_{\odot}} \right)^t e^{u\mathrm{[Fe/H]}} \, ,
\label{eq:scal_freq_meso}
\end{equation}
with $\beta$ once again a scaling factor and $s, t, u$ the exponents that need to be estimated. For this property of the meso-granulation we consider the linearized form of Eq.~(\ref{eq:scal_freq_meso}), and test four different models, which we label as $\mathcal{M}_{b,1}$ for $t = 0$ and $u = 0$, $\mathcal{M}_{b,2}$ for $u = 0$, $\mathcal{M}_{b,3}$ for $t = 0$, and $\mathcal{M}_{b,4}$ for $t \ne 0$ and $u \ne 0$. In this formulation, model $\mathcal{M}_{b,1}$ is clearly the simplest, not including both mass and metallicity terms, while models $\mathcal{M}_{b,2}$ and $\mathcal{M}_{b,3}$ consider the dependence on mass and metallicity separately from one another and are equally complex in terms of parameters. Model $\mathcal{M}_{b,4}$ is instead the most generalized one, where both mass and metallicity effects are included at the same time.

Following the same approach used for the meso-granulation amplitude, we derive the analytical expressions for the parameter-dependent relative uncertainties associated with the predicted meso-granulation frequencies, yielding the general form
\begin{equation}
\widetilde{\sigma}^2_{b} (s,t,u) = \widetilde{\sigma}_{b_\mathrm{meso}}^2 + s^2 \widetilde{\sigma}_{\nu_\mathrm{max}}^2 + t^2 \widetilde{\sigma}_M^2 + u^2 \widetilde{\sigma}_{\rm{[Fe/H]}}^2 \, ,
\end{equation}
with the same definitions as in Sect.~\ref{sec:scal_amp} for $\nu_\mathrm{max}$, $M$, and [Fe/H], and with $\widetilde{\sigma}_{b_\mathrm{meso}} \equiv \sigma_{b_\mathrm{meso}}/b_\mathrm{meso}$. Like for the amplitudes, we can obtain the total relative uncertainty for the frequency predictions by imposing $t = 0$ and $u = 0$ for model $\mathcal{M}_{b,1}$, $u = 0$ for model $\mathcal{M}_{b,2}$, $t = 0$ for model $\mathcal{M}_{b,3}$, and $t \ne 0$ and $u \ne 0$ for model $\mathcal{M}_{b,4}$.

\section{Bayesian inference}
\label{sec:inference}
We perform a Bayesian inference on the models presented in Sect.~\ref{sec:scal_rel} by adopting a Gaussian likelihood where the residuals, assumed to be Gaussian distributed, arise from the difference between the observed and predicted natural logarithms of the parameters that describe the granulation activity. The Gaussian log-likelihood, similar to \cite{Corsaro13} and \cite{Bonanno14}, therefore reads
\begin{equation}
\Lambda (\boldsymbol{\theta}) = \Lambda_0 (\boldsymbol{\theta}) - \frac{1}{2} \sum^N_{i=1} \left[ \frac{\Delta_i (\boldsymbol{\theta})}{\widetilde{\sigma_i} (\boldsymbol{\theta})} \right]^2 \, ,
\label{eq:likelihood}
\end{equation}
where $\boldsymbol{\theta}$ is the parameter vector, for example $(\beta,s,t, u)$ for the model $\mathcal{M}_{a,2}$, $N$ is the total number of stars, and $\Lambda_0 (\boldsymbol{\theta})$ is a term depending on the relative uncertainties, given by
\begin{equation}
\Lambda_0 (\boldsymbol{\theta}) = - \sum^N_{i=1} \ln \sqrt{2 \pi} \widetilde{\sigma_i} (\boldsymbol{\theta}) \, .
\end{equation}
The residuals between observed and predicted values are defined as
\begin{equation}
\Delta_i (\boldsymbol{\theta}) = \ln a_\mathrm{meso}^\mathrm{obs} - \ln a_\mathrm{meso}^\mathrm{pred} (\boldsymbol{\theta}) \, .
\end{equation}
The results from the Bayesian parameter estimation are listed in Table~\ref{tab:param_gran}, with $\Lambda_\mathrm{max}$ representing the maximum value of the log-likelihood function (Eq.~\ref{eq:likelihood}), increasing as the fit to the data improves. We note that in order to evaluate whether the fit quality of a model is better than that of other models, $\Lambda_\mathrm{max}$ has to be compared to that of a competitor model and has therefore no meaning on its own. A thorough assessment of the statistical significance of a model is presented in Sect.~\ref{sec:model_comparison}. The predictions obtained from the estimated parameters are compared to the observations in Fig.~\ref{fig:fit_amp} and \ref{fig:fit_freq} for all the models considered in Sect~\ref{sec:scal_rel}.

\begin{table*}
\centering
 \caption{Median values of the inferred parameters $(s, t, u, \ln \beta)$ for all the models presented in Sect.~\ref{sec:scal_rel}, with the physical parameter they relate to shown in brackets. Bayesian credible intervals of 68.3\,\% are added. The maximum value for the log-likelihood function, $\Lambda_\mathrm{max}$, is reported as fit quality indicator, where a larger value corresponds to a better fit to the observations.}
\begin{tabular}{lrrrrr}
\hline\hline
\\[-8pt]
Model & \multicolumn{1}{c}{$s\,(\numax)$} & \multicolumn{1}{c}{$t\,(M)$} & \multicolumn{1}{c}{$u$ ([Fe/H])} & \multicolumn{1}{c}{$\ln \beta$} & $\Lambda_\mathrm{max}$\\[1pt]
 \hline
 \\[-8pt]       
  $\mathcal{M}_{a,1}$ & $-0.550^{+0.008}_{-0.009}$ & $-0.67^{+0.02}_{-0.02}$ & \multicolumn{1}{c}{--} & $-0.25^{+0.04}_{-0.04}$ & $-46.4$\\ [1pt]	%  LN(Bayesian Evidence): -56.02062244
  $\mathcal{M}_{a,2}$ & $-0.593^{+0.010}_{-0.010}$ & $-0.21^{+0.04}_{-0.05}$ & $0.89^{+0.08}_{-0.08}$ & $-0.71^{+0.06}_{-0.06}$ & $-31.3$\\ [1pt]	%  LN(Bayesian Evidence): -41.60166030	
  \hline
 \\[-8pt]        
  $\mathcal{M}_{b,1}$ & $0.954^{+0.007}_{-0.008}$ & \multicolumn{1}{c}{--} & \multicolumn{1}{c}{--} & $0.03^{+0.03}_{-0.03}$ & $-227.4$\\[1pt]		%  LN(Bayesian Evidence): -237.29763497
  $\mathcal{M}_{b,2}$ & $0.917^{+0.008}_{-0.009}$ & $0.20^{+0.02}_{-0.02}$ & \multicolumn{1}{c}{--} & $-0.19^{+0.03}_{-0.03}$ & $-120.2$\\[1pt]	%  LN(Bayesian Evidence): -129.27513818
  $\mathcal{M}_{b,3}$ & $0.889^{+0.009}_{-0.009}$ & \multicolumn{1}{c}{--} & $-0.52^{+0.03}_{-0.03}$ & $-0.16^{+0.03}_{-0.03}$ & $-37.1$\\[1pt]		%  LN(Bayesian Evidence): -45.36780603
  $\mathcal{M}_{b,4}$ & $0.898^{+0.012}_{-0.014}$ & $-0.38^{+0.06}_{-0.06}$ & $-1.15^{+0.12}_{-0.10}$ & $0.10^{+0.06}_{-0.07}$ & $-2.5$\\[1pt]		%  LN(Bayesian Evidence): -12.59555190
  \hline
 \end{tabular}
\label{tab:param_gran}
\end{table*}

\subsection{Model hypothesis testing}
\label{sec:model_comparison}
The Bayesian model hypothesis test is performed by computing the so-called odds ratio between two competing models $\mathcal{M}_i$ and $\mathcal{M}_j$
\begin{equation}
\mathcal{O}_{ij} = \frac{\mathcal{E}_{i}}{\mathcal{E}_{j}} \frac{\pi ( \mathcal{M}_i )}{\pi ( \mathcal{M}_j )} = \mathcal{B}_{ij} \frac{\pi ( \mathcal{M}_i )}{\pi ( \mathcal{M}_j )} \, ,
\label{eq:odds_ratio}
\end{equation}
where $\mathcal{B}_{ij}$ is the Bayes factor given as the ratio of the Bayesian evidences ($\mathcal{E}$) of the two models, and $\pi (\mathcal{M})$ is our model prior, or equivalently model weight, assigned to each of the models investigated. Given the linearity of the models, model priors for multiplicity adjustment can be taken into account \citep{Scott10}. For this purpose, we consider the model prior function proposed by \cite{Scott10}, which for a model having $k$ free parameters out of a full set of $m$ free parameters investigated, i.\,e. the total number of parameters to test, reads as
\begin{equation}
\pi (\mathcal{M}^k) = \frac{k! (m-k)!}{m! (m+1)} \, .
\end{equation}
In our analysis, the linearized models for amplitudes (see Sect.~\ref{sec:scal_amp}) and characteristic frequency (see Sect.~\ref{sec:scal_freq}) account for a total of $m=3$ free parameters $(s, t, u)$ related to the observables. The intercept $\ln \beta$ is not included in the count of free parameters relevant for the model prior because a model with the intercept as the only free parameter is the null model, with $k = 0$. The model priors give $\pi(\mathcal{M}^\mathrm{k=3}) / \pi(\mathcal{M}^\mathrm{k=2}) = \pi(\mathcal{M}^\mathrm{k=3}) / \pi(\mathcal{M}^\mathrm{k=1}) = 3$. 

We compute the Bayes factor for each pair of scaling relations following \cite{Corsaro13}. Since the Bayesian evidence of a model taken singularly is not meaningful, the best (or statistically more likely) model is chosen as the one that maximizes the odds ratio given by Eq.~(\ref{eq:odds_ratio}) in a comparison between pairs of models, for all the models considered in the analysis. Results for the model comparison for all the models investigated in this work are presented in Table~\ref{tab:bayes_gran} for $a_\mathrm{meso}$ and $b_\mathrm{meso}$, with the most favored models highlighted with shades. The net effects caused by stellar mass and metallicity on the meso-granulation properties can be isolated by adopting the most favored scaling relations selected by our model comparison process. The results are depicted in Figs.~\ref{fig:amp_numax}b and \ref{fig:freq_numax}b, both as a function of $\numax$, and in Figs.~\ref{fig:meso_mass} and \ref{fig:meso_feh} as a function of mass and metallicity, respectively.

\begin{table}
\centering
 \caption{Natural logarithms of the odds ratio, $\ln \mathcal{O}_{ij} = \ln \pi(\mathcal{M}_i) - \ln \pi (\mathcal{M}_j) + \ln \mathcal{E}_i - \ln \mathcal{E}_j $, for each pair of models $(\mathcal{M}_i$, $\mathcal{M}_j$ related to the meso-granulation amplitude $a_\mathrm{meso}$ (marked with subscript a), and characteristic frequency $b_\mathrm{meso}$ (marked with subscript b). Shading indicates the most likely models ($\mathcal{M}_{a,2}$ for $a_\mathrm{meso}$ and $\mathcal{M}_{b,4}$ for $b_\mathrm{meso}$), meaning that they maximize all the odds ratios when compared to the competitor models.}
\begin{tabular}{lrrrrr}
\hline\hline
\\[-8pt]
Model & \multicolumn{1}{c}{$\mathcal{M}_{a,1}$} & \multicolumn{1}{c}{$\mathcal{M}_{a,2}$} & \multicolumn{1}{c}{$\mathcal{M}_{b,1}$} & \multicolumn{1}{c}{$\mathcal{M}_{b,2}$} & \multicolumn{1}{c}{$\mathcal{M}_{b,3}$}\\[1pt]
 \hline       
\\[-8pt]
  $\mathcal{M}_{a,1}$ & \multicolumn{1}{c}{--}  & \\ [2pt]
\cellcolor[gray]{0.9}$\mathcal{M}_{a,2}$ & \cellcolor[gray]{0.9}$15.5$ & \multicolumn{1}{c}{--} \\ [1pt]
\hline
\\[-8pt]
$\mathcal{M}_{b,1}$ & & & \multicolumn{1}{c}{--}  & & \\ [2pt]
$\mathcal{M}_{b,2}$ & & & $108.0$ & \multicolumn{1}{c}{--} & \\ [2pt]
$\mathcal{M}_{b,3}$ & & & $191.9$ & $83.9$ & \multicolumn{1}{c}{--} \\ [2pt]
\cellcolor[gray]{0.9}$\mathcal{M}_{b,4}$ & \cellcolor[gray]{0.9} & \cellcolor[gray]{0.9} &  \cellcolor[gray]{0.9}$225.8$ & \cellcolor[gray]{0.9}$117.8$ & \cellcolor[gray]{0.9}$33.9$\\ [1pt]
  \hline
 \end{tabular}
\label{tab:bayes_gran}
\end{table}

\section{Results}
\label{sec:results}
From our inference on the models presented in Sect.~\ref{sec:scal_rel}, stellar mass and especially metallicity appear to play a statistically significant role in the meso-granulation properties of the stars. In Figs.~\ref{fig:amp_numax}a and \ref{fig:amp_numax}c, the latter showing the amplitudes after the mass effect has been removed, we observe two distinct groups of stars. The first group corresponds to stars with super-solar metallicity and high meso-granulation amplitudes (from NGC\,6791), while the second one is composed of stars with close-to-solar metallicity and low meso-granulation amplitudes (from NGC\,6819 and NGC\,6811). We observe these two groups independently of whether a star is RC or RGB. The trend with metallicity is also clearly shown in the top panel of Fig.~\ref{fig:meso_feh}, where the amplitudes have been corrected for $\numax$ and mass dependencies. 

In Figs.~\ref{fig:freq_numax}a and \ref{fig:freq_numax}c, the latter showing meso-granulation frequencies without the mass effect, we see that the stars belonging to NGC\,6791 have the tendency to exhibit frequencies smaller than the stars of the other two clusters. This is more evident from the bottom panel of Fig.~\ref{fig:meso_feh}, in which the meso-granulation frequencies were rescaled to remove the effect of a varying $\numax$ and stellar mass. These observational considerations are reflected in the values of the exponents of the scaling relations and in our model hypothesis testing, which we discuss in Sect.~\ref{sec:results_agran} and Sect.~\ref{sec:results_bgran}. 

From both Fig.~\ref{fig:amp_numax}a and Fig.~\ref{fig:freq_numax}a we note that the typical $\nu_\mathrm{max}$ of the RC stars within a cluster changes significantly from one cluster to another, with $\nu_\mathrm{max} \sim 20\,\mu$Hz for NGC\,6791, $\nu_\mathrm{max} \sim 35\,\mu$Hz for NGC\,6819, and $\nu_\mathrm{max} \sim 100\,\mu$Hz for NGC\,6811. This difference is mainly caused by the different average masses of the RC stars in each cluster (Fig.~\ref{fig:teff_mass}) because for a constant stellar radius $\nu_\mathrm{max} \propto M T_\mathrm{eff}^{-0.5}$ (see also the color-magnitude diagram shown in Fig.~\ref{fig:cmd} and those presented in Fig. 9 of \citealt{Corsaro12}). A detailed discussion of our findings can be found in Sect.~\ref{sec:discussion}, where we highlight their implications and physical interpretations.

\subsection{Meso-granulation amplitude $a_\mathrm{meso}$}
\label{sec:results_agran}
We find that the scaling relations for $a_\mathrm{meso}$ have a negative exponent $t$, for stellar mass, of $-0.67$ and $-0.21$ (models $\mathcal{M}_{a,1}$ and $\mathcal{M}_{a,2}$ respectively), and a positive exponent $u$, for [Fe/H], of about $0.89$ (model $\mathcal{M}_{a,2}$). The results from the parameter estimation of the best model $\mathcal{M}_{a,2}$ imply that the dependency on metallicity is more than four times stronger than that on stellar mass, and that an increasing metallicity increases the amplitude of the signal (see Fig.~\ref{fig:amp_numax}c and our fit marked in Fig.~\ref{fig:meso_feh}, top panel). If we consider stars at constant surface gravity --- to first approximation at constant $\numax$ given that the temperature range of our sample of stars is not large ($\sim10^3$\,K) --- the effect of metallicity is opposite to that of stellar mass (Fig.~\ref{fig:amp_numax}d versus Fig.~\ref{fig:amp_numax}c, and Fig.~\ref{fig:meso_mass} versus Fig.~\ref{fig:meso_feh}, top panels). 

The dependency on metallicity estimated from the best model $\mathcal{M}_{a,2}$ is about 1.5 times stronger than the dependency on $\nu_\mathrm{max}$, hence than on $g / \sqrt{T_\mathrm{eff}}$ and to first approximation on surface gravity, with an exponent $s$ set around $-0.55$ and $-0.59$ for models $\mathcal{M}_{a,1}$ and $\mathcal{M}_{a,2}$, respectively. The exponent $s$ is close, although not compatible within the quoted errors, to the value of $-0.61$ found by K14 using a larger sample of field stars that also included main sequence stars. The exponent $t$ of model $\mathcal{M}_{a,2}$, including [Fe/H], is also compatible (within 1-$\sigma$) with that found by K14, $-0.26$, while model $\mathcal{M}_{a,1}$ has a three times larger estimate of $t$ with respect to that of model $\mathcal{M}_{a,2}$. The stronger dependency on stellar mass in the scaling relation associated to model $\mathcal{M}_{a,1}$, as compared to that of $\mathcal{M}_{a,2}$, is also a consequence of the lack of a term that takes into account the different stellar metallicity, which is significantly different in NGC\,6791 with respect to NGC\,6811 and NGC\,6811 (a factor of about two). Differences between our exponents and those from the literature also rely on: (i) the adoption of a sample of only RGs (K14 included both field RGs and main sequence stars), hence of a range of surface gravities ($2.3 \leq \log g \leq 3.1$) and temperatures ($4350$\,K $< T_\mathrm{eff} < 5150$\,K) typical of evolved low- and intermediate-mass stars; (ii) the use of different data sources and of corrected stellar masses (as derived in Sect.~\ref{sec:mass}); (iii) the use of a more accurate and uniform set of fundamental stellar properties, stemming from the cluster membership of the targets. 

In regard to the solar reference values used in this work (Sect.~\ref{sec:virgo}), our estimation of the proportionality term $\beta$ suggests that the reference amplitude for our sample should be smaller than $a_\mathrm{meso,\odot}$, by $\sim22$\,\% and $\sim50$\,\% for models $\mathcal{M}_{a,1}$ and $\mathcal{M}_{a,2}$, respectively. This probably signals a break-down of linearity across the orders of magnitude in surface gravity, that separate our sample of RGs from the Sun.

As shown in Table~\ref{tab:bayes_gran}, the Bayesian model comparison largely favors model $\mathcal{M}_{a,2}$, including the metallicity term, against $\mathcal{M}_{a,1}$ because the corresponding odds ratio ($\ln \mathcal{O}_{2,1} \simeq 16$) is well above a strong evidence condition ($\ln \mathcal{O}_{ij} > 5$ for model $\mathcal{M}_i$ versus model $\mathcal{M}_j$, according to the Jeffreys' scale of strength), thus justifying the inclusion of an additional dependency on metallicity. This is also observed in the much higher maximum likelihood value of $\mathcal{M}_{a,2}$ compared to that of $\mathcal{M}_{a,1}$, with $\Lambda_\mathrm{max}^{a,2} - \Lambda_\mathrm{max}^{a,1} \simeq 15$ (see Table~\ref{tab:param_gran}). We therefore recommend the adoption of the scaling relation given by Eq.~(\ref{eq:scal_amp_meso}) when predicting meso-granulation amplitudes for RGs having $20\,\mu$Hz $< \numax < 160\,\mu$Hz. Although for solar metallicity stars, [Fe/H]\,$ = 0$, model  $\mathcal{M}_{a,2}$ formally reduces to the analytical form of $\mathcal{M}_{a,1}$, the exponent for the stellar metallicity is $u \neq 0$ and for the exponents $s$ (for $\numax$) and $t$ (for stellar mass) of the scaling relation estimates from model $\mathcal{M}_{a,2}$, as listed in Table~\ref{tab:param_gran}, should still be taken into account. Finally, as seen from Fig.~\ref{fig:fit_amp}, and also apparent in Fig.~\ref{fig:amp_numax}b where we show the amplitudes corrected for mass and metallicity effects, we do not observe any clear difference between RC and RGB stars because the residuals from our predictions are on the same level for both evolutionary stages (on average around $8$\,\%).

\subsection{Meso-granulation characteristic frequency $b_\mathrm{meso}$}
\label{sec:results_bgran}
For the characteristic frequency of the meso-granulation we have tested the four different models described in Sect.~\ref{sec:scal_freq}. In this case, an increasing metallicity appears to reduce $b_\mathrm{meso}$, namely to increase the time scale of the meso-granulation. This can be seen from Fig.~\ref{fig:freq_numax}a, and more so from Fig.~\ref{fig:freq_numax}c, where the mass effect has been removed, and from the bottom panel of Fig.~\ref{fig:meso_feh}, in which the trend with metallicity has been isolated from the strong dependency on $\numax$. This result is confirmed by the exponents estimated for Eq.~(\ref{eq:scal_freq_meso}). We find that the exponent related to metallicity, $u$, is $-0.52$ for model $\mathcal{M}_{b,3}$ and $-1.15$ for the best model $\mathcal{M}_{b,4}$, indicating that the strength of the relation between $b_\mathrm{meso}$ and [Fe/H] is comparable to that between $b_\mathrm{meso}$ and $\nu_\mathrm{max}$. 

An exponent $s \simeq 0.9$, which is found for all the models for $b_\mathrm{meso}$ tested in this work, shows that $b_\mathrm{meso}$ and $\nu_\mathrm{max}$ are almost linearly related, implying that the two parameters do follow a similar scaling (see Sect.~\ref{sec:discussion} for a discussion on this result). The $s$ exponent found in this work is on average only 5\% smaller, although significant according to the quoted errors, than that obtained by K14 using a similar scaling relation for the same meso-granulation component. Once again we attribute this difference to the different sample, range of fundamental stellar properties, and sources of observational data used.

According to our estimates of the proportionality term $\beta$, we find that an optimal reference value would range from $\sim83$\,\% (model $\mathcal{M}_{b,2}$), up to $\sim110$\,\% of $b_\mathrm{meso,\odot}$ (model $\mathcal{M}_{b,4}$). Overall, this is closer to the adopted solar value than what was found in the case of the meso-granulation amplitude, and the difference from unity in the parameter $\beta$ is not even statistically significant for model $\mathcal{M}_{b,1}$, and only marginally significant for model $\mathcal{M}_{b,4}$ (within $2$-$\sigma$). This suggests that a possible break-down in linearity for the characteristic timescale of the granulation and meso-granulation signals between our sample and the Sun is in general less likely than for amplitudes.

The effect of mass on $b_\mathrm{meso}$, after surface gravity has been accounted for, is in the same direction as that of metallicity (Fig.~\ref{fig:freq_numax}d versus Fig.~\ref{fig:freq_numax}c and Fig.~\ref{fig:meso_mass} versus Fig.~\ref{fig:meso_feh}, bottom panels). The associated exponent $t$ reaches up to only $-0.38$ for the best model $\mathcal{M}_{b,4}$. This weak mass dependence is also evident from the Bayesian model comparison (Table~\ref{tab:bayes_gran}) where model $\mathcal{M}_{b,3}$, incorporating only metallicity and $\numax$ dependencies, far exceeds the strong evidence condition against model $\mathcal{M}_{b,2}$ ($\ln \mathcal{O}_{3,2} \simeq 84$), the latter including only stellar mass and $\numax$. However, $\mathcal{M}_{b,4}$, which encompasses both mass and [Fe/H] on top of $\numax$, is significantly better than a model that incorporates one or the other ($\ln \mathcal{O}_{4,2} \simeq 118$, $\ln \mathcal{O}_{4,3} \simeq 34$). We note that the odds ratio between model $\mathcal{M}_{b,4}$ and model $\mathcal{M}_{b,3}$ is much smaller than that between $\mathcal{M}_{b,3}$ and $\mathcal{M}_{b,2}$, confirming that even in this case the stellar mass does not constitute a dominant contribution to $b_\mathrm{meso}$. Like $a_\mathrm{meso}$, we recommend the adoption of Eq.~(\ref{eq:scal_freq_meso}) for predictions of the meso-granulation characteristic frequency for RGs in the $\numax$ range investigated. We show the fit results in Fig.~\ref{fig:fit_freq}, where the scatter in the residuals is on average around $8$-$9$\,\% for the different models, and in Fig.~\ref{fig:freq_numax}b where we correct for mass and metallicity effects. Similar to the case of $a_\mathrm{meso}$, we do not find any evidence for a systematic difference between RC and RGB stars of the same cluster.

\subsection{Assessing the reliability of the metallicity effect}
\label{sec:reliability}
To further validate our results we perform three additional analyses described below. First, to make sure no biases are caused by the numerical method implemented in \diamonds, we derive the background parameters discussed in Sect.~\ref{sec:bkg} using another automated fitting routine, based on a Bayesian maximum a posteriori method \citep{Gaulme09}. We find that the resulting values of $\nu_\mathrm{max}$, $a_\mathrm{meso}$, and $b_\mathrm{meso}$ agree with the measurements derived with \diamonds\,\,on average within 1, 3, and 6\,\%, respectively, compatible within the Bayesian credible intervals. 

\begin{figure}[htb]
   \centering
   \includegraphics[width=9.0cm]{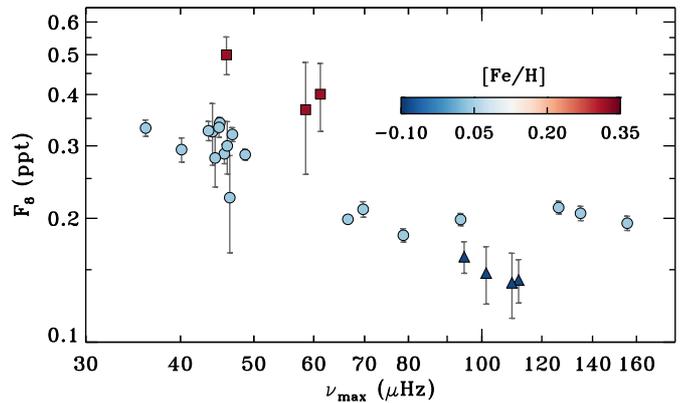}
      \caption{Granulation flicker for timescales shorter than 8~hr, $F_8$, as a function of $\nu_\mathrm{max}$ for NGC\,6791 (squares), NGC\,6819 (circles) and NGC\,6811 (triangles). The color-coding shows the cluster mean metallicity, similarly to Fig.~\ref{fig:amp_numax}c, panel c. $1$-$\sigma$ measurement uncertainties on flicker are also shown.}
    \label{fig:flicker_numax}
\end{figure}

Second, we measure the granulation flicker, $F_8$, introduced by \cite{Bastien13Nature}, from the \kepler light curves for the stars in our sample satisfying the limits of applicability defined in \citet{Bastien16} (a total of 26 targets, 3 from NGC\,6819, 4 from NGC\,6811, and 19 from NGC\,6819). For each of these stars we therefore have the amplitude in parts-per-thousand (ppt) of the total granulation signal on timescales shorter than 8~hr. The granulation flicker represents a measurement of the granulation activity that is independent of the background modeling adopted in Sect.~\ref{sec:bkg}. The result is shown in Fig.~\ref{fig:flicker_numax} as a function of $\nu_\mathrm{max}$ and [Fe/H]. To test the significance of the dependency on metallicity on top of those arising from a varying stellar mass and $\numax$, we consider a scaling relation of the form
\begin{equation}
F_8 = \alpha \left( \frac{\nu_\mathrm{max}}{\nu_\mathrm{max,\odot}} \right)^s \left( \frac{M}{M_{\odot}} \right)^t e^{u\mathrm{[Fe/H]}} \, ,
\end{equation}
with $\alpha$ a proportionality term in units of ppt, and $s, t, u$ exponents that need to be estimated. We thus apply the same Bayesian inference described in Sect.~\ref{sec:inference} to both the linearized models determined for $u = 0$ (no metallicity effect) and $u \neq 0$ (metallicity included). The Bayesian model comparison between the two models considered, performed as described in Sect.~\ref{sec:model_comparison}, shows that the model including the metallicity term is significantly dominant over its competitor accounting only for $\numax$ and stellar mass ($\ln \mathcal{O}_{u \neq 0, u = 0} = 5.2$). From our granulation flicker we find a metallicity exponent, $u=0.9\pm 0.3$, which agrees, well within the $1$-$\sigma$ error, with that estimated for the meso-granulation amplitude from our detailed analysis of the background spectra presented in Sect.~\ref{sec:results_agran}. This positive detection of the metallicity effect was possible despite the granulation flicker could be measured for only three of the high-metallicity stars of NGC\,6791. We note that the uncertainties on $F_8$, and the errors on the parameters estimated from the fit to the whole sample, are about three times larger than for the analysis of the background spectra.

Lastly, we measure the granulation properties $a_\mathrm{meso}$ and $b_\mathrm{meso}$, together with $\nu_\mathrm{max}$, for an independent sample of 12 field RGs. These field stars have temperatures and metallicities available from the APOKASC catalog and all exhibit similar apparent magnitudes, which implies that the noise level in the stellar PSD is similar from star to star. This homogeneity in apparent magnitude, unlike our cluster RGs, is useful to isolate possible effects in estimating the background parameters that could arise from a different noise level in the data. The field stars are divided into two groups, the first one including six targets with solar metallicity (simulating the metallic content of the stars in NGC\,6819 and NGC 6811), and the second group with six targets having super-solar metallicity, on average $\sim0.24$\,dex, hence simulating a sample of stars with a metallic content close to that of NGC\,6791. For the selected field stars we considered masses obtained from scaling using the asteroseismic parameters from the APOKASC DR13 catalog \citep{Pinsonneault14APOKASC}, and a $\nu_\mathrm{max}$ that is estimated through \diamonds\,\,using the same background model presented in Sect.~\ref{sec:bkg}. Each star of a set with similar metallicity has a relatively close comparative star in terms of mass and $\nu_\mathrm{max}$ in the other set of super-solar metallicity targets. Our choice of masses and $\nu_\mathrm{max}$ values for the field RGs allows us to soften the effect caused by a different stellar mass at constant surface gravity and to limit the tight dependence on $g/\sqrt{T_\mathrm{eff}}$, which contributes each time that two targets having different metallicities are compared to one another. We observe a trend with metallicity for both meso-granulation amplitude and characteristic frequency similar to the one shown in this work. In particular, by considering the predictions obtained from the best meso-granulation amplitude scaling relation, Eq.~(\ref{eq:scal_amp_meso}), we obtain average residuals of $\sim$28\,ppm and $\sim$24\,ppm for the super-solar and solar metallicity samples, respectively. These estimates are similar to the dispersion of the residuals for the same scaling relation applied to the cluster RGs (about 28\,ppm, see also Fig.~\ref{fig:fit_amp}). The comparison done for the best meso-granulation frequency scaling relation, Eq.~(\ref{eq:scal_freq_meso}), shows that the average residuals are about 3.3\,$\mu$Hz and 3.5\,$\mu$Hz for the super-solar and the solar metallicity samples, respectively. Similarly to the case of the amplitudes, this result is again close to the dispersion of the residuals found in the cluster sample (about 2\,ppm, see Fig.\ref{fig:fit_freq}). We note that a dedicated and detailed analysis of the granulation activity for field stars that aims at calibrating the granulation scaling relations for a wide range of stellar parameters will be presented by Mathur et al. (in prep.). We therefore conclude that the systematic difference in amplitude and frequency of the meso-granulation signal that we observe in Fig.~\ref{fig:amp_numax} and \ref{fig:freq_numax} could not be caused by either the different signal-to-noise ratio in the PSDs of the cluster RGs or just the differing stellar masses.

\section{Discussion \& conclusions}
\label{sec:discussion}
From the results presented in Sect.~\ref{sec:results} it appears clearly that the stellar metallicity has an important influence on the granulation activity in evolved cool stars (Fig.~\ref{fig:meso_feh}) and that meso-granulation and granulation properties depend solely on the conditions in the stellar atmospheres. We have shown that the cluster membership of the 60 RGs analyzed in this work is a powerful constraint that allows accurate calibration of the meso-granulation scaling relations in the metallicity range spanning from 0.8 times to about twice the solar metallicity ($\sim$0.4\,dex), and surface gravity $2.3 \leq \log g \leq 3.1$. This is because the stars in each cluster have the advantage of sharing rather homogeneous stellar properties of mass, temperature, and metallicity (see Appendix~\ref{sec:appendix_a} and Table~\ref{tab:cluster_feh}). The results of this work are therefore essential to identify and understand the underlying correlations among metallicity and stellar mass, surface gravity, and temperature. This study also sets the basis for the selection and detailed analysis of large samples of field stars spanning a wider range of fundamental stellar properties and evolutionary stages than the one covered here (Mathur et al. in prep).

The signature of metallicity is statistically significant for both the amplitude and the characteristic frequency of the meso-granulation, hence of the granulation since the two components scale linearly across the entire $\numax$ range investigated (see Sect.~\ref{sec:bkg}). From our sample, the effect of metallicity is observationally more enhanced for the amplitude parameter, for which we clearly observe a systematic difference between the close-to-solar metallicity stars and the super-solar metallicity ones (Fig.~\ref{fig:amp_numax}a, c). This is because metallicity has the opposite effect on the amplitudes than that of mass (Figs.~\ref{fig:meso_mass} and \ref{fig:meso_feh}, top panels), and because at the same time the high-metallicity cluster, NGC\,6791, contains the stars with the lowest mass among the three clusters considered. In particular, $a_\mathrm{meso}$ for stars belonging to NGC\,6791 is about 60\,\% larger than that of stars from NGC\,6819 and NGC\,6811, and this result is a combination of both metallicity and mass of the stars. These findings are also reflected in our measurements of the granulation flicker, $F_8$, where the high-metallicity stars show systematically larger flicker amplitudes than the low-metallicity ones (Fig.~\ref{fig:flicker_numax}), and with a similar dependence on metallicity as estimated from the associated exponent $u$ of the scaling relations (see Sect.~\ref{sec:reliability}). Unlike for the amplitudes, mass and metallicity act in the same direction for the characteristic frequency $b_\mathrm{meso}$, namely by decreasing the frequency scale of the meso-granulation, hence of the granulation, when they increase (Figs.~\ref{fig:meso_mass} and \ref{fig:meso_feh}, bottom panels). This also explains that the metallicity effect seen in Fig.~\ref{fig:freq_numax}a is less evident than that of the amplitudes, but it otherwise appears clear when the mass dependency is removed, as seen in Fig.~\ref{fig:freq_numax}c and in Fig.~\ref{fig:meso_feh}, bottom panel.

\subsection{Meso-granulation amplitude $a_\mathrm{meso}$}
\label{sec:discussion_agran}
According to our analysis of the scaling relations presented in Sect.~\ref{sec:scal_rel}, a higher metallicity increases the granulation amplitude. The best model identified ($\mathcal{M}_{a,2}$, Eq.~\ref{eq:scal_amp_meso}, see Table~\ref{tab:param_gran}) gives a power law exponent for metallicity of $u = 0.89$, with a precision of about $9$\,\%. The positive value of the exponent is in qualitative agreement with theoretical modeling of realistic 3D stellar atmospheres by \cite{Collet07granulation}. In particular, from the modeled granulation larger granules are found for higher metallicities due to the increased opacity. This implies that the amplitudes of the meso-granulation and granulation signals are higher because the associated disk-integrated brightness fluctuations scale as $n_\mathrm{gran}^{-1/2}$, $n_\mathrm{gran}$ being the number of granules observed on the stellar surface \citep{Ludwig06gran}. When the granulation scale is fixed, that is, when the atmospheric parameters are fixed, $n_{\rm gran}$ scales with stellar radius as $R^{2}$ and the amplitude of the granulation signal therefore scales as $R^{-1}$, due to, for example, a change in stellar mass. The metallicity effect is also apparent in the new granulation analysis by \cite{Ludwig16}, of 3D surface convection simulations of F-K stars, from the main sequence to the lower RGB and for solar and metal-poor, [Fe/H]$=-2$, compositions. They Fourier transformed time-series of specific bolometric intensity, integrated over the disk, and scaled to a star of solar radius to make it easy to apply their result to stars of any radius. For our sample of giants, the difference between bolometric and \kepler intensities will be rather small. From \cite{Ludwig16} Fig.~3, the increase in granulation amplitude with increasing metallicity is more pronounced toward larger values of surface gravity (or equivalently larger $\nu_\mathrm{max}$), which we also find in our observations (Fig.~\ref{fig:amp_numax}a,c, in the range $60-120\,\mu$Hz). 

We perform a direct comparison between our observed granulation amplitudes and the predictions from the 3D simulations by \cite{Ludwig16}. We compare with the solar metallicity stars of our sample, from NGC\,6819 and NGC\,6811. For these stars we reduce the \kepler meso-granulation amplitudes to bolometric \emph{granulation} amplitudes. We further reduce these amplitudes to that of a star with solar radius (having the same $T_{\rm eff}$ and $\log g$) by multiplying by $(R_\mathrm{*}/R_\odot)$. The stellar radii are computed according to \cite{Sharma16} (see also Appendix A). The bi-linear fits to the scaled bolometric amplitudes of \cite{Ludwig16}, for both [Fe/H] $= 0$ and $-2$ regimes are then applied to the $T_\mathrm{eff}$ (from the SDSS-based temperature scale) and $\log g$ of all the targets of NGC\,6819 and 6811 (see Tables~\ref{tab:stellar_parameters_NGC6819} and \ref{tab:stellar_parameters_NGC6811}) to obtain predicted estimates from the 3D simulations. The simulations from \cite{Ludwig16} exhibit $0.27$\,dex ($+85$\%) larger amplitudes for a $2.0$\,dex increase in metallicity in the range of temperature and surface gravities covered by our sample of RGs. By scaling the change in logarithmic amplitude down to the $0.32$\,dex increase in metallicity of our targets (from the two solar metallicity clusters, to NGC\,6791) gives a $0.045$\,dex ($+11$\%) increase in amplitude. This agrees well with our observations, for which we find a $0.050$\,dex ($+12$\%) increase in amplitude between the clusters. Interestingly, our best amplitude model, $\mathcal{M}_{a,2}$, is more sensitive to metallicity than $\nu_\mathrm{max}$ ($|s| < |u|$, see Table~\ref{tab:param_gran}). 

As noted by K14, by exploiting the linear relation between $b_\mathrm{gran}$ and $\numax$, and by evaluating the background PSD at $\numax$ from our Eq.~(\ref{eq:overall_bkg}), we can infer that $a_\mathrm{gran} \propto \numax^{-0.5}$. We can easily extend this relation to the meso-granulation amplitude since both $a_\mathrm{meso}$ and $b_\mathrm{meso}$ follow constant scalings from $a_\mathrm{gran}$ and $b_\mathrm{gran}$ (see, e.g., K14, and our discussions in Sect.~\ref{sec:bkg} and Sect.~\ref{sec:discussion_bgran}). Our estimate of the exponent $s$ of the scaling relations for $a_\mathrm{meso}$ is  close to $-0.5$ (with $-0.550$ for $\mathcal{M}_{a,1}$ and $-0.593$ for $\mathcal{M}_{a,2}$), thus in better agreement with the expected dependency than what was found in previous works (see, e.g., K14). 

According to our scaling relations we find that the effect of a varying mass at constant surface gravity on $a_\mathrm{meso}$, could be well constrained from the population of cluster RGs thanks to the homogeneity of stellar masses found within each cluster (see Fig.~\ref{fig:teff_mass}, Fig.~\ref{fig:amp_numax}d, and Fig.~\ref{fig:meso_mass}, top panel). By taking into account RGs that exhibit, on average, a different mass depending on the cluster they belong to (see also \citealt{Miglio12massloss,Corsaro12} for more details), the effect of mass can be clearly disentangled from that of a different surface gravity and metallicity. The statistical error for the mass exponent, $t$, is about 19\,\%, comparable to that obtained by K14 using 100 times as many field stars but about one year shorter time-series. From the parameter estimation presented in Table~\ref{tab:param_gran}, it appears that the impact of a varying mass on $a_\mathrm{meso}$ is about a fourth in strength as compared to that of metallicity (for $\mathcal{M}_{a,2}$ we have $4 | t | \simeq | u |$). We note that since the amplitude of the signal scales with stellar radius as $R^{-1}$, for a constant surface gravity the amplitude scales as $M^{-1/2}$, which is not far from the estimates of the mass exponent $t$ for both models $\mathcal{M}_{a,1}$ and $\mathcal{M}_{a,2}$ ($t = -0.67$ and $-0.21$, respectively). Deviations from the expected exponent in mass are most likely due to temperature, surface gravity, and metallicity dependencies that are not entirely described by a $\numax$ dependency.

Our analysis also has some implications on the connection between the granulation properties and the stellar evolutionary stage. On one hand, as visible from the residuals of the fits shown in Fig.~\ref{fig:fit_amp}, we do not observe any systematic differences between RC and RGB stars. This confirms that the meso-granulation, hence also the granulation, are unaffected by the conditions in the stellar core because they are completely described by the atmospheric parameters and by the stellar radius, which produces an attenuation to the global power of granular fluctuations by a factor $R^{-2}$ due to the stochastic and incoherent nature of granulation \citep{Trampedach13gran}. On the other hand, the coefficients $\beta$ estimated for the two meso-granulation amplitude models, and especially the best model $\mathcal{M}_{a,2}$, suggest that the scaling relations might not be linear across the orders of magnitudes in surface gravity that separate the Sun from the evolved stars in our sample.

From our estimation of the background parameters shown in Fig.~\ref{fig:amp_numax}a, we notice two stars that have meso-granulation amplitudes significantly above than of similar stars, and significantly above our predictions from the scaling relations (by up to 25\,\% as seen in Fig.~\ref{fig:fit_amp}). These are the RGB star KIC\,2437976, with $\nu_\mathrm{max} \simeq 89\,\mu$Hz, and the RC star KIC\,2437103, with $\numax \simeq 29\,\mu$Hz, both in NGC\,6791. Given the accuracy of the $\nu_\mathrm{max}$ and $\Delta\nu$ measurements for the stars, and their cluster membership, we discard errors in the mass computation as the source of the discrepancy. KIC\,2437103 in particular, was originally misclassified from its position in the color-magnitude diagram as an RGB star, but then reclassified as RC from the properties of its oscillation \citep[see][for a detailed discussion]{Corsaro12}. This star has the largest $a_\mathrm{meso}$ of our sample, exceeding 700\,ppm (see Table~\ref{tab:bkg_parameters_NGC6791}), which can in part be interpreted as a consequence of the combination of the low stellar mass, one of the smallest among all the stars that we analyzed (see also Fig.~\ref{fig:amp_numax}d), and the low $\numax$. We conclude that the large amplitudes observed for these two stars could be the result of either a metallicity higher than the cluster mean metallicity adopted in our study, a possible blending with other sources (e.g., from a binary), or a combination of the two. Future abundance determinations, which are not yet available for the two mentioned stars, might help understanding the origin of the observed discrepancy. 

\subsection{Meso-granulation characteristic frequency $b_\mathrm{meso}$}
\label{sec:discussion_bgran}
By analyzing four different scaling relations we have found that metallicity has a more dominant role than mass in determining the characteristic frequency of meso-granulation. Stellar mass has a rather weak effect, with an exponent $t = -0.38$, as compared to that of metallicity (from our best model $\mathcal{M}_{b,4}$, Eq.~\ref{eq:scal_freq_meso}, we have $|u| \simeq 3 |t|$), although still statistically significant (see Table~\ref{tab:bayes_gran}). Once again, despite the relatively low number of stars in our sample as compared to that of K14, the metallicity exponent $u$ could be constrained to about $10$\,\%, while the precision remains poorer on the mass exponent $t$ (about $16$\,\%). In contrast to the meso-granulation amplitudes, we do not find a clear correspondence with the modeling performed by \cite{Ludwig16}. This is because in the metallicity range explored by the authors ($-2 \leq$ [Fe/H] $\leq 0$) the granulation characteristic frequency appears to have an almost negligible dependency upon metallicity, with variations of $\sim6$\,\%, comparable to the level of the residuals found from the scaling relations investigated in this work. Interestingly, for our best model we find that $|u| \simeq 1.3 |s|$, indicating that $\nu_\mathrm{max}$ and [Fe/H] are almost equally important in determining $b_\mathrm{meso}$. We conclude that in order to assess our results on the characteristic frequency of the granulation activity, more dedicated theoretical investigations would be required, for example with 3D simulations covering the atmospheric parameters of our sample of cluster stars, and by extracting granulation parameters from the Fourier spectra of time-series obtained from such simulations, based on the background fitting models used in Sect.~\ref{sec:bkg}. Mathur et al. (in prep.) is investigating the metallicity effect, based on \kepler observations of field stars, and the convection simulations by \cite{Trampedach13gran} (see \citealt{Mathur11granulation} for the granulation parameters for this grid) to cover a wider range of metallicity, surface gravities, and temperatures, than that of \cite{Ludwig16}.

The exponent $s$, of $\nu_\mathrm{max}$, is $0.898$ for our best model $\mathcal{M}_{b,4}$ (incorporating both metallicity and mass in the fit) and shows that the relation between $b_\mathrm{meso}$ and $\nu_\mathrm{max}$ is nearly linear as a first approximation, thus validating the presence of a tight connection between the granulation characteristic frequency and the surface gravity of the star. As argued by \cite{Kjeldsen11} and by \cite{Mathur11granulation} (see also K14), this confirms that convection cells travel for a vertical distance that is proportional to the pressure scale height, $H_p$, at a speed proportional to the speed of sound $c_\mathrm{s}$ to first approximation. For explaining this result, we consider that the derived characteristic frequency of the granulation can be expressed as $b_\mathrm{gran} \propto c_s / H_p$. Taking into account the seismic scaling relation $\numax \propto g/\sqrt{T_\mathrm{eff}}$ \citep{Brown91numax}, and the relations $c_\mathrm{s} \propto \sqrt{T_\mathrm{eff}}$ and $H_p \propto T_\mathrm{eff}/g$, we thus have that $b_\mathrm{gran} \propto \numax$ because it is the same convection that excites the acoustic oscillations and gives rise to the granulation at the surface. This in turn implies that $b_\mathrm{meso} \propto \numax$, because the meso-granulation represents conglomerations of a certain number of granules, hence meso-granulation and granulation have time scales that are proportional to each other, $b_\mathrm{meso} \propto b_\mathrm{gran}$, as we also quantify in Sect.~\ref{sec:bkg}.

Like $a_\mathrm{meso}$, the distribution of $b_\mathrm{meso}$ for RGB and RC stars is similar (e.g., see Fig.~\ref{fig:fit_freq}). Once again this indicates that granulation and meso-granulation inherently depend on the atmospheric conditions only, with no effect from the different core structures between RGB and RC stages of stellar evolution. In addition, the power of the granulation signal decreases with stellar radius due to the averaging over an increasing number of incoherent granules. This manifests in our fits from the best model $\mathcal{M}_\mathrm{b,4}$, as a negative mass-exponent $t$.

\begin{acknowledgements}
E.C. is funded by the European Union's Horizon 2020 research and innovation program under the Marie Sklodowska-Curie grant agreement n$^\circ$ 664931 and by the European Community's Seventh Framework Programme (FP7/2007-2013) under grant agreement n$^\circ$312844 (SPACEINN). S.M. acknowledges support from NASA grants NNX12AE17G and NNX15AF13G and NSF grant AST-1411685. R.A.G. received funding from the CNES GOLF and PLATO grants at CEA and from the ANR (Agence Nationale de la Recherche, France) program IDEE (n$^\circ$ ANR-12-BS05-0008) ``Interaction Des \'Etoiles et des Exoplan\`etes''.
\\\\
Funding for the Sloan Digital Sky Survey IV has been provided by the
Alfred P. Sloan Foundation, the U.S. Department of Energy Office of
Science, and the Participating Institutions. SDSS acknowledges
support and resources from the Center for High-Performance Computing at
the University of Utah. The SDSS web site is www.sdss.org.
\\\\
SDSS is managed by the Astrophysical Research Consortium for the Participating Institutions of the SDSS Collaboration including the Brazilian Participation Group, the Carnegie Institution for Science, Carnegie Mellon University, the Chilean Participation Group, the French Participation Group, Harvard-Smithsonian Center for Astrophysics, Instituto de Astrof\'{i}sica de Canarias, The Johns Hopkins University, Kavli Institute for the Physics and Mathematics of the Universe (IPMU) / University of Tokyo, Lawrence Berkeley National Laboratory, Leibniz Institut f\"{u}r Astrophysik Potsdam (AIP), Max-Planck-Institut f\"{u}r Astronomie (MPIA Heidelberg), Max-Planck-Institut f\"{u}r Astrophysik (MPA Garching), Max-Planck-Institut f\"{u}r Extraterrestrische Physik (MPE), National Astronomical Observatories of China, New Mexico State University, New York University, University of Notre Dame, Observat\'{o}rio Nacional / MCTI, The Ohio State University, Pennsylvania State University, Shanghai Astronomical Observatory, United Kingdom Participation Group, Universidad Nacional Aut\'{o}noma de M\'{e}xico, University of Arizona, University of Colorado Boulder, University of Oxford, University of Portsmouth, University of Utah, University of Virginia, University of Washington, University of Wisconsin, Vanderbilt University, and Yale University.
\end{acknowledgements}

\bibliographystyle{aa} % style aa.bst

\appendix
\section{Stellar atmospheric parameters and masses}
\label{sec:appendix_a}
Our selection of stellar atmospheric parameters from the different sources presented in Sect.~\ref{sec:data} is shown in Tables~\ref{tab:stellar_parameters_NGC6791}, \ref{tab:stellar_parameters_NGC6819}, \ref{tab:stellar_parameters_NGC6811}, for NGC\,6791, NGC\,6819, and NGC\,6811, respectively. The values for surface gravity, $\log g$, are also provided for completeness and are derived from corrected stellar masses and radii computed according to \cite{Sharma16}. The 1-$\sigma$ uncertainty in $\log g$ is determined from a standard propagation from those of corrected masses and radii.

\begin{table}
\tiny
\caption{Atmospheric parameters $T_\mathrm{eff}$ (from SDSS-based temperature scale, see Sect.~\ref{sec:temperature}, with a total uncertainty of 69\,K for all stars that are not marked by an apex), [Fe/H] from ASPCAP where available, and corrected stellar masses for NGC 6791 following \cite{Sharma16}. Surface gravities, $\log g$, and corresponding 1-$\sigma$ uncertainties, are also provided using stellar radii computed according to \cite{Sharma16}. The evolutionary state specifies whether a star is RC or RGB, as classified by \cite{Corsaro12,Corsaro17}.}             % title of Table
\centering
\begin{tabular}{l l r c c l}       
\hline\hline
\\[-8pt]
KIC ID & \multicolumn{1}{c}{$T_\mathrm{eff}$} & \multicolumn{1}{c}{[Fe/H]} & Mass & $\log g$ & State\\ [1pt]
 & \multicolumn{1}{c}{(K)} & & ($M_{\odot}$) & (dex) & \\ [1pt]
 \hline             
 2297384$^{a}$  &  $4504$  &  $0.38  \pm  0.02$ &  $ 1.06       \pm 0.04         $ &  $      2.37      \pm       0.02       $ &	RC	\\ [1pt]	 
 2297825$^{a}$  &  $4479$  &  $0.31  \pm  0.02$ &  $ 0.95       \pm 0.05         $ &	 $      2.37      \pm       0.02    $ & RC	\\ [1pt]
 2435987  &  $4427$  &  $0.28  \pm  0.02$ &		$ 1.09       \pm 0.03         $ & $      2.46      \pm       0.01       $ & RGB 	\\ [1pt]
 2436097  &  $4402$  & \multicolumn{1}{c}{--} & 	$ 1.11       \pm 0.03         $ & $      2.51      \pm       0.02       $ & RGB	\\ [1pt]
 2436417  &  $4460$  &   $0.31  \pm  0.02$ & 		$ 1.04       \pm 0.04         $ & $      2.32      \pm       0.02       $ & RC 	\\ [1pt]
 2436458  &  $4350$  &  \multicolumn{1}{c}{--} & 	$ 1.05       \pm 0.03         $ & $      2.45      \pm       0.01       $ & RGB 	\\ [1pt]
 2436676  &  $4573$  &  \multicolumn{1}{c}{--} & 	$ 0.97       \pm 0.03         $ & $      3.01      \pm       0.01       $ & RGB	\\ [1pt]
 2436732  &  $4503$  &  \multicolumn{1}{c}{--} & 	$ 1.10       \pm 0.03         $ & $      2.37      \pm       0.02       $ & RC	\\ [1pt]
 2436818  &  $4610$  &  \multicolumn{1}{c}{--} & 	$ 1.04       \pm 0.03         $ & $      2.87      \pm       0.01       $ & RGB	\\ [1pt]
 2437103  &  $4503$  &  \multicolumn{1}{c}{--} & 	$ 0.83       \pm 0.03         $ & $      2.35      \pm       0.02       $ & RC	\\ [1pt]
 2437240  &  $4459$  &  \multicolumn{1}{c}{--} & 	$ 1.15       \pm 0.03         $ & $      2.55      \pm       0.01       $ & RGB	\\ [1pt]
 2437270$^{a}$  &  $4499$ &  \multicolumn{1}{c}{--} &  $ 1.26       \pm 0.05         $ & $      2.73      \pm       0.02       $ & RGB	\\ [1pt]
 2437325  &  $4484$  &  \multicolumn{1}{c}{--} & 	$ 1.04       \pm 0.03         $ & $      2.86      \pm       0.01       $ & RGB	\\ [1pt]
 2437353  &  $4520$ &  $0.30  \pm  0.02$ & 		$ 1.10       \pm 0.03         $ & $      2.39      \pm       0.01       $ &  RC	\\ [1pt]
 2437564  &  $4467$  &  $0.32  \pm  0.02$ & 		$ 1.12       \pm 0.03         $ & $      2.39      \pm       0.02       $ & RC	\\ [1pt]
 2437589$^{a}$  &  $4508$  & \multicolumn{1}{c}{--} &  $ 1.44       \pm 0.06         $ & $      2.55      \pm       0.02       $ & RGB	\\ [1pt]
 2437804  &  $4439$  &  $0.35  \pm  0.03$ & 		$ 1.02       \pm 0.03         $ & $      2.31      \pm       0.02       $ & RC	\\ [1pt]
 2437933  &  $4534$  & \multicolumn{1}{c}{--} & 	$ 1.11       \pm 0.03         $ & $      2.92      \pm       0.01       $ & RGB	\\ [1pt]
 2437957  &  $4556$  & \multicolumn{1}{c}{--} & 	$ 1.03       \pm 0.03         $ & $      2.86      \pm       0.01       $ & RGB	\\ [1pt]
 2437972$^{a}$  & $4543$  & \multicolumn{1}{c}{--} &  $ 1.09       \pm 0.04         $ & $      2.82      \pm       0.02       $ & RGB	\\ [1pt]
 2437976  &  $4478$  & \multicolumn{1}{c}{--} & 	$ 1.05       \pm 0.03         $ & $      2.84      \pm       0.01       $ & RGB	\\ [1pt]
 2437987  &  $4517$  & \multicolumn{1}{c}{--} & 	$ 1.13       \pm 0.05         $ & $      2.38      \pm       0.02       $ & RC	\\ [1pt]
 2438038$^{a}$  &  $4450$  & \multicolumn{1}{c}{--} &  $ 1.09       \pm 0.04         $ & $      2.68      \pm       0.02       $ & RGB\\ [1pt]
 2438051  &  $4524$  &  $0.30  \pm  0.02$ & 		$ 1.14       \pm 0.07         $ & $      2.37      \pm       0.03       $ & RC \\ [1pt]
 2438333  &  $4473$  &  $0.32  \pm  0.02$ & 		$ 1.08       \pm 0.03         $ & $      2.67      \pm       0.01       $ & RGB \\ [1pt]
 2569055  &  $4479$  &  $0.32  \pm  0.02$ & 		$ 1.09       \pm 0.04         $ & $      2.38      \pm       0.02       $ & RC \\ [1pt]
 2569945  &  $4507$  &  $0.33  \pm  0.02$ & 		$ 0.97       \pm 0.04         $ & $      2.37      \pm       0.02       $ & RC \\ [1pt]
 2570094$^{a}$  &  $4485$  & \multicolumn{1}{c}{--} &  $ 1.16       \pm 0.04         $ & $      2.72      \pm       0.02       $ & RGB\\ [1pt]
 2570244  &  $4515$  & \multicolumn{1}{c}{--} & 	$ 1.09       \pm 0.03         $ & $      2.91      \pm       0.01       $ & RGB \\ [1pt]
 2570384  &  $4519$  &  $0.32  \pm  0.02$ & 		$ 0.81       \pm 0.05         $ & $     2.65      \pm       0.03       $ & RGB \\ [1pt]
\hline
\\[-8pt]
%\\[1pt]                
\end{tabular}
\tablefoot{$^{a}$ The source for temperature is the $(V-K)$ color \citep{Hekker11OC}, with an adopted uncertainty of 110\,K. See Sect.~\ref{sec:temperature} for more details.}
\label{tab:stellar_parameters_NGC6791}
\end{table}

\begin{table}
\tiny
\caption{Same as in Table~\ref{tab:stellar_parameters_NGC6791} but for NGC 6819.}             % title of Table
\centering                         
\begin{tabular}{l l r c c l}       
\hline\hline
\\[-8pt]
KIC ID & \multicolumn{1}{c}{$T_\mathrm{eff}$} & \multicolumn{1}{c}{[Fe/H]} & Mass & $\log g$ & State\\ [1pt]
 & \multicolumn{1}{c}{(K)} & & ($M_{\odot}$) & (dex) & \\ [1pt]
 \hline          
4937056    & $4844$  &  $0.01 \pm 0.02$ &  $ 1.56       \pm 0.07         $ 	&	$      2.57      \pm       0.02       $ & RC\\ [1pt]
4937770    & $5033$  &  $-0.02 \pm 0.02$ & $ 1.93       \pm 0.08         $ 	&	$      2.88      \pm       0.02       $ & RC\\ [1pt]
5023953    & $4834$  &  $0.07 \pm 0.02$ & $ 1.84       \pm 0.05         $	&	$      2.59      \pm       0.01       $ & RC\\ [1pt]
5024327    & $4865$  &  \multicolumn{1}{c}{--} & $ 1.50       \pm 0.05         $ 	&	$      2.55      \pm       0.02       $ & RC\\ [1pt]
5024404    & $4798$  &  $0.07 \pm 0.02$ & $ 1.52       \pm 0.04         $	&	$      2.57      \pm       0.01       $ & RC\\ [1pt]
5024414    & $5031$  &  \multicolumn{1}{c}{--} & $ 2.40       \pm 0.05         $ 	&	$      2.81      \pm       0.01       $ & RC\\ [1pt]
5024476    & $4968$  &  $0.05 \pm 0.02$ & $ 2.21       \pm 0.05         $ 	&	$      2.73      \pm       0.01       $ & RC\\ [1pt]
5024582    & $4873$  &  \multicolumn{1}{c}{--} & $ 1.54       \pm 0.04         $ 	&	$      2.57      \pm       0.01       $ & RC\\ [1pt]
5024967    & $4797$  &  \multicolumn{1}{c}{--} & $ 1.52       \pm 0.05         $ &	$      2.56      \pm       0.02       $ & RC\\ [1pt]
5111718    & $4916$  &  $0.08 \pm   0.02$ & $ 1.60       \pm 0.04         $ 	&	$      3.04      \pm       0.01       $ & RGB\\ [1pt]
5111949    & $4804$  &  $0.07 \pm   0.03$ & $ 1.49       \pm 0.04         $ 	&	$      2.57      \pm       0.01       $ & RC\\ [1pt]
5112072    & $4937$  &  $0.01 \pm   0.03$ & $ 1.60       \pm 0.03         $ &	 $      3.01      \pm       0.01       $ &RGB\\ [1pt]
5112361    & $4924$  &  $-0.03 \pm   0.03$ & $ 1.87       \pm 0.04         $ &	$      2.75      \pm       0.01       $ & RGB\\ [1pt]
5112373    & $4826$  &  $0.04 \pm   0.03$ & $ 1.51       \pm 0.04         $ &	 $      2.55      \pm       0.01       $ & RC\\ [1pt]
5112387    & $4808$  &  $0.05 \pm   0.02$ & $ 1.49       \pm 0.04         $ &	 $      2.55      \pm       0.01       $ & RC\\ [1pt]
5112401    & $4797$  &  $0.01 \pm   0.03$ & $ 1.32       \pm 0.04         $ &	 $      2.46      \pm       0.02       $ & RC\\ [1pt]
5112467    & $4841$  &  $0.05 \pm   0.03$ & $ 1.51       \pm 0.04         $ &	 $      2.57      \pm       0.01       $ & RC\\ [1pt]
5112491    & $4894$  &  $0.00 \pm   0.02$ & $ 1.50       \pm 0.04         $ &	 $      2.55      \pm       0.01       $ & RC\\ [1pt]
5112730    & $4794$  &  $0.06 \pm   0.02$ & $ 1.54       \pm 0.04         $ &	 $      2.54      \pm       0.01       $ & RC\\ [1pt]
5112938    & $4798$  &  $0.06 \pm   0.02$ & $ 1.41       \pm 0.04         $ &	 $      2.55      \pm       0.01       $ & RC\\ [1pt]
5112950    & $4746$  &  $0.07 \pm   0.02$ & $ 1.51       \pm 0.04         $ &	 $      2.51      \pm       0.02       $ & RC\\ [1pt]
5112974    & $4790$  &  $0.03 \pm   0.02$ & $ 1.50       \pm 0.04         $ 	& $      2.50      \pm       0.01       $ &	RC\\ [1pt]
5113441    & $4829$  &  $0.08  \pm  0.02$ & $ 1.55       \pm 0.03         $ 	& $      3.09      \pm       0.01       $ &	RGB\\ [1pt]
5200152    & $4927$  &  $0.05  \pm  0.02$ & $ 1.51       \pm 0.04         $ 	&	$      2.56      \pm       0.02       $ & RC\\ [1pt]
\hline
\\[-8pt]
%\\[1pt]                
\end{tabular}
\label{tab:stellar_parameters_NGC6819}
\end{table}

\begin{table}
\tiny
\caption{Same as in Table~\ref{tab:stellar_parameters_NGC6791} but for NGC 6811.}             % title of Table
\centering                         
\begin{tabular}{l l r c c l}       
\hline\hline
\\[-8pt]
KIC ID & \multicolumn{1}{c}{$T_\mathrm{eff}$} & \multicolumn{1}{c}{[Fe/H]} & Mass & $\log g$ & State\\ [1pt]
 & \multicolumn{1}{c}{(K)} & & ($M_{\odot}$) & (dex) &\\ [1pt]
 \hline
 9532903  &  $5106$ & $-0.06 \pm 0.02$ &  $ 2.27 \pm 0.06 $ & $      2.89      \pm       0.01       $ & RC\\ [1pt]
9534041  &  $5144$ & $-0.11 \pm 0.02$ & $ 2.72 \pm 0.06 $ & $      2.96      \pm       0.01       $ & RC\\ [1pt]
9655101  &  $5067$ & $-0.11 \pm 0.03$ & $ 2.54 \pm 0.06 $ & $      2.92      \pm       0.01       $ & RC\\ [1pt]
9716090  &  $5084$ & \multicolumn{1}{c}{--} & $ 2.25 \pm 0.05 $ & $      2.95      \pm       0.01       $ & RC\\ [1pt]
9716522  &  $4985$ & $-0.06 \pm 0.03$ & $ 2.47 \pm 0.06 $ & $      2.64      \pm       0.01       $ & RC\\ [1pt]
9776739  &  $5152$ & \multicolumn{1}{c}{--}& $ 2.27 \pm 0.06 $ & $      2.89      \pm       0.01       $ & RC\\ [1pt]
\hline
\\[-8pt]
%\\[1pt]                
\end{tabular}
\tablefoot{Sources for temperature and metallicities are as in Table~\ref{tab:stellar_parameters_NGC6791}. The evolutionary state of the stars KIC~9776739 and KIC~9716090 is provided by \cite{Molenda14NGC6811} and confirmed by Corsaro et la. (in prep.).}
\label{tab:stellar_parameters_NGC6811}
\end{table}

\section{Results for the background fitting}
\label{sec:bkg_results}
The background parameters $a_\mathrm{meso}$, $b_\mathrm{meso}$, $\nu_\mathrm{max}$ derived by means of \diamonds\,\,for the entire sample of 60 cluster RGs analyzed in this work are listed for \ngca\,\,in Table~\ref{tab:bkg_parameters_NGC6791}, \ngcb\,\,in Table~\ref{tab:bkg_parameters_NGC6819}, and \ngcc\,\,in Table~\ref{tab:bkg_parameters_NGC6811}.

Following the definitions presented by \cite{Corsaro14}, the configuring parameters used in \diamonds\,\,are: initial enlargement fraction $f_0 = 1.3$, shrinking rate $\alpha = 0.02$, number of live points $N_\mathrm{live} = 500$, number of clusters $1 \leq N_\mathrm{clust} \leq 10$, number of total drawing attempts $M_\mathrm{attempts} = 10^4$, number of nested iterations before the first clustering $M_\mathrm{init} = 1500$, and number of nested iterations with the same clustering $M_\mathrm{same} = 50$.

\begin{table}
\caption{Median values with corresponding 68.3\,\% Bayesian credible intervals of the background parameters $a_\mathrm{gran}$, $b_\mathrm{gran}$, $\numax$ for the RGs of the open cluster NGC\,6791, as derived using \diamonds.}             % title of Table
\centering
\begin{tabular}{l l r r}       
\hline\hline
\\[-8pt]         
KIC ID & \multicolumn{1}{c}{$a_\mathrm{meso}$} & \multicolumn{1}{c}{$b_\mathrm{meso}$} & \multicolumn{1}{c}{$\numax$}\\ [1pt]
 & \multicolumn{1}{c}{(ppm)}  & \multicolumn{1}{c}{(\muhz)} & \multicolumn{1}{c}{(\muhz)}\\ [1pt]   
\hline
\\[-8pt]
2297384  & $     561.9       ^{+      17.4      }_{-      16.5   } $ & $      10.5       ^{+       0.5      }_{-       0.4   } $ & $      30.6       ^{+       0.2      }_{-       0.2} $\\[1pt]
2297825  & $     583.8       ^{+      19.0      }_{-      18.3   } $ & $      10.4       ^{+       0.5      }_{-       0.5   } $ & $      30.4       ^{+       0.3      }_{-       0.3} $\\[1pt]
2435987  & $     477.8       ^{+       9.9      }_{-       9.9   } $ & $      10.4       ^{+       0.4      }_{-       0.4   } $ & $      37.8       ^{+       0.2      }_{-       0.2} $\\[1pt]
2436097  & $     465.7       ^{+      11.1      }_{-      12.0   } $ & $      12.4       ^{+       0.6      }_{-       0.6   } $ & $      42.1       ^{+       0.3      }_{-       0.2} $\\[1pt]
2436417  & $     604.0       ^{+      28.5      }_{-      25.6   } $ & $      10.1       ^{+       0.7      }_{-       0.7   } $ & $      27.4       ^{+       0.2      }_{-       0.2} $\\[1pt]
2436458  & $     499.2       ^{+      10.2      }_{-      10.3   } $ & $      10.8       ^{+       0.4      }_{-       0.4   } $ & $      37.1       ^{+       0.2      }_{-       0.2} $\\[1pt]
2436676  & $     223.8       ^{+       6.8      }_{-       6.9   } $ & $      35.8       ^{+       1.0      }_{-       1.0   } $ & $     131.5       ^{+       0.6      }_{-       0.7} $\\[1pt]
2436732  & $     580.6       ^{+      12.6      }_{-      10.9   } $ & $       8.9       ^{+       0.3      }_{-       0.3   } $ & $      30.3       ^{+       0.2      }_{-       0.2} $\\[1pt]
2436818  & $     276.2       ^{+       5.1      }_{-       4.8   } $ & $      26.4       ^{+       0.6      }_{-       0.5   } $ & $      95.5       ^{+       0.5      }_{-       0.5} $\\[1pt]
2437103  & $     711.3       ^{+      20.8      }_{-      21.5   } $ & $       8.1       ^{+       0.5      }_{-       0.4   } $ & $      28.8       ^{+       0.2      }_{-       0.2} $\\[1pt]
2437240  & $     448.3       ^{+       9.1      }_{-       9.7   } $ & $      12.9       ^{+       0.4      }_{-       0.5   } $ & $      46.0       ^{+       0.2      }_{-       0.2} $\\[1pt]
2437270  & $     311.5       ^{+       6.3      }_{-       5.0   } $ & $      16.2       ^{+       0.5      }_{-       0.5   } $ & $      69.9       ^{+       0.2      }_{-       0.2} $\\[1pt]
2437325  & $     276.9       ^{+       5.8      }_{-       6.4   } $ & $      23.3       ^{+       0.4      }_{-       0.4   } $ & $      94.1       ^{+       0.2      }_{-       0.2} $\\[1pt]
2437353  & $     508.9       ^{+      10.3      }_{-      11.8   } $ & $       9.1       ^{+       0.4      }_{-       0.3   } $ & $      31.7       ^{+       0.2      }_{-       0.2} $\\[1pt]
2437564  & $     537.8       ^{+      15.4      }_{-      17.3   } $ & $       9.9       ^{+       0.5      }_{-       0.5   } $ & $      32.0       ^{+       0.2      }_{-       0.2} $\\[1pt]
2437589  & $     364.0       ^{+       7.5      }_{-       8.2   } $ & $      14.5       ^{+       0.5      }_{-       0.5   } $ & $      46.1       ^{+       0.3      }_{-       0.3} $\\[1pt]
2437804  & $     612.6       ^{+      16.3      }_{-      14.9   } $ & $       8.1       ^{+       0.4      }_{-       0.3   } $ & $      26.7       ^{+       0.2      }_{-       0.2} $\\[1pt]
2437933  & $     262.4       ^{+       6.8      }_{-       5.9   } $ & $      28.9       ^{+       1.0      }_{-       0.9   } $ & $     108.4       ^{+       0.3      }_{-       0.3} $\\[1pt]
2437957  & $     279.2       ^{+      22.5      }_{-      14.6   } $ & $      28.4       ^{+       1.2      }_{-       1.1   } $ & $      92.7       ^{+       0.3      }_{-       0.4} $\\[1pt]
2437972  & $     358.8       ^{+      28.4      }_{-      21.5   } $ & $      19.8       ^{+       0.9      }_{-       1.0   } $ & $      85.2       ^{+       0.3      }_{-       0.3} $\\[1pt]
2437976  & $     412.7       ^{+      24.8      }_{-      22.3   } $ & $      26.2       ^{+       1.7      }_{-       1.5   } $ & $      89.3       ^{+       0.4      }_{-       0.3} $\\[1pt]
2437987  & $     576.5       ^{+      18.7      }_{-      19.6   } $ & $       8.5       ^{+       0.5      }_{-       0.6   } $ & $      31.0       ^{+       0.4      }_{-       0.4} $\\[1pt]
2438038  & $     351.7       ^{+       6.8      }_{-       7.1   } $ & $      15.3       ^{+       0.5      }_{-       0.5   } $ & $      62.5       ^{+       0.2      }_{-       0.2} $\\[1pt]
2438051  & $     590.3       ^{+      15.8      }_{-      18.2   } $ & $       9.3       ^{+       0.4      }_{-       0.4   } $ & $      30.1       ^{+       0.6      }_{-       0.5} $\\[1pt]
2438333  & $     372.1       ^{+       7.0      }_{-       6.2   } $ & $      15.1       ^{+       0.4      }_{-       0.4   } $ & $      61.2       ^{+       0.2      }_{-       0.2} $\\[1pt]
2569055  & $     564.8       ^{+      16.9      }_{-      16.5   } $ & $       9.5       ^{+       0.5      }_{-       0.5   } $ & $      31.0       ^{+       0.3      }_{-       0.3} $\\[1pt]
2569945  & $     551.0       ^{+      19.4      }_{-      18.9   } $ & $      10.3       ^{+       0.6      }_{-       0.5   } $ & $      30.4       ^{+       0.4      }_{-       0.4} $\\[1pt]
2570094  & $     324.2       ^{+       7.6      }_{-       7.8   } $ & $      16.0       ^{+       0.5      }_{-       0.5   } $ & $      68.1       ^{+       0.2      }_{-       0.2} $\\[1pt]
2570244  & $     245.2       ^{+       5.9      }_{-       5.8   } $ & $      24.0       ^{+       0.7      }_{-       1.0   } $ & $     105.8       ^{+       0.4      }_{-       0.4} $\\[1pt]
2570384  & $     410.9       ^{+      28.4      }_{-      19.7   } $ & $      19.1       ^{+       1.7      }_{-       1.4   } $ & $      58.5       ^{+       0.9      }_{-       1.1} $\\[1pt]
\hline                                
\end{tabular}
\label{tab:bkg_parameters_NGC6791}
\end{table}

\begin{table}
\caption{Same as in Table~\ref{tab:bkg_parameters_NGC6791} but for NGC\,6819.}             % title of Table
\centering                         
\begin{tabular}{l l r r}       
\hline\hline
\\[-8pt]         
KIC ID & \multicolumn{1}{c}{$a_\mathrm{meso}$} & \multicolumn{1}{c}{$b_\mathrm{meso}$} & \multicolumn{1}{c}{$\numax$}\\ [1pt]
 & \multicolumn{1}{c}{(ppm)}  & \multicolumn{1}{c}{(\muhz)} & \multicolumn{1}{c}{(\muhz)}\\ [1pt]   
\hline
\\[-8pt]
4937056  & $     344.8       ^{+       7.8      }_{-       7.4   } $ & $      14.1       ^{+       0.6      }_{-       0.6   } $ & $      46.3       ^{+       0.7      }_{-       0.6} $\\[1pt]
4937770  & $     162.9       ^{+      10.2      }_{-       8.7   } $ & $      32.9       ^{+       2.1      }_{-       2.9   } $ & $      93.8       ^{+       1.1      }_{-       1.0} $\\[1pt]
5023953  & $     301.9       ^{+       3.9      }_{-       3.7   } $ & $      14.1       ^{+       0.3      }_{-       0.3   } $ & $      48.7       ^{+       0.2      }_{-       0.2} $\\[1pt]
5024327  & $     344.0       ^{+       8.4      }_{-       8.4   } $ & $      15.5       ^{+       0.6      }_{-       0.5   } $ & $      44.4       ^{+       0.4      }_{-       0.3} $\\[1pt]
5024404  & $     325.7       ^{+       4.6      }_{-       4.7   } $ & $      12.6       ^{+       0.3      }_{-       0.3   } $ & $      47.0       ^{+       0.2      }_{-       0.2} $\\[1pt]
5024414  & $     180.4       ^{+       3.2      }_{-       3.2   } $ & $      27.5       ^{+       1.6      }_{-       1.5   } $ & $      78.8       ^{+       0.2      }_{-       0.2} $\\[1pt]
5024476  & $     189.0       ^{+       4.2      }_{-       4.0   } $ & $      20.5       ^{+       0.4      }_{-       0.4   } $ & $      66.6       ^{+       0.3      }_{-       0.3} $\\[1pt]
5024582  & $     311.2       ^{+       4.2      }_{-       4.1   } $ & $      13.6       ^{+       0.3      }_{-       0.3   } $ & $      46.5       ^{+       0.2      }_{-       0.2} $\\[1pt]
5024967  & $     320.5       ^{+       6.3      }_{-       7.1   } $ & $      14.7       ^{+       0.5      }_{-       0.6   } $ & $      45.7       ^{+       0.4      }_{-       0.4} $\\[1pt]
5111718  & $     172.2       ^{+       1.9      }_{-       1.8   } $ & $      36.5       ^{+       0.6      }_{-       0.8   } $ & $     135.0       ^{+       0.3      }_{-       0.3} $\\[1pt]
5111949  & $     351.3       ^{+       5.0      }_{-       5.6   } $ & $      14.6       ^{+       0.4      }_{-       0.4   } $ & $      46.8       ^{+       0.2      }_{-       0.2} $\\[1pt]
5112072  & $     173.1       ^{+       2.6      }_{-       2.9   } $ & $      35.1       ^{+       0.4      }_{-       0.6   } $ & $     126.3       ^{+       0.2      }_{-       0.2} $\\[1pt]
5112361  & $     211.9       ^{+       3.5      }_{-       3.0   } $ & $      21.4       ^{+       0.5      }_{-       0.5   } $ & $      69.7       ^{+       0.2      }_{-       0.2} $\\[1pt]
5112373  & $     347.0       ^{+       5.8      }_{-       6.4   } $ & $      14.6       ^{+       0.4      }_{-       0.4   } $ & $      44.1       ^{+       0.2      }_{-       0.2} $\\[1pt]
5112387  & $     333.4       ^{+       6.0      }_{-       6.9   } $ & $      14.3       ^{+       0.4      }_{-       0.5   } $ & $      45.1       ^{+       0.2      }_{-       0.2} $\\[1pt]
5112401  & $     409.3       ^{+       8.8      }_{-       7.8   } $ & $      14.0       ^{+       0.5      }_{-       0.4   } $ & $      36.0       ^{+       0.2      }_{-       0.3} $\\[1pt]
5112467  & $     308.8       ^{+       5.4      }_{-       5.1   } $ & $      14.4       ^{+       0.5      }_{-       0.4   } $ & $      46.1       ^{+       0.2      }_{-       0.2} $\\[1pt]
5112491  & $     310.6       ^{+       4.9      }_{-       5.3   } $ & $      14.3       ^{+       0.4      }_{-       0.4   } $ & $      44.4       ^{+       0.2      }_{-       0.2} $\\[1pt]
5112730  & $     342.4       ^{+       7.4      }_{-       6.7   } $ & $      13.9       ^{+       0.6      }_{-       0.6   } $ & $      43.6       ^{+       0.2      }_{-       0.2} $\\[1pt]
5112938  & $     348.0       ^{+       5.8      }_{-       5.3   } $ & $      13.9       ^{+       0.3      }_{-       0.4   } $ & $      45.0       ^{+       0.2      }_{-       0.2} $\\[1pt]
5112950  & $     352.3       ^{+       6.9      }_{-       7.5   } $ & $      13.1       ^{+       0.5      }_{-       0.6   } $ & $      41.3       ^{+       0.3      }_{-       0.2} $\\[1pt]
5112974  & $     329.1       ^{+       6.4      }_{-       6.4   } $ & $      13.0       ^{+       0.4      }_{-       0.5   } $ & $      40.1       ^{+       0.2      }_{-       0.2} $\\[1pt]
5113441  & $     170.0       ^{+       0.4      }_{-       0.3   } $ & $      47.0       ^{+       0.4      }_{-       0.4   } $ & $     155.6       ^{+       0.1      }_{-       0.1} $\\[1pt]
5200152  & $     355.4       ^{+       7.0      }_{-       8.4   } $ & $      14.8       ^{+       0.5      }_{-       0.5   } $ & $      45.1       ^{+       0.3      }_{-       0.3} $\\[1pt]                                
\hline
\end{tabular}
\label{tab:bkg_parameters_NGC6819}
\end{table}

\begin{table}
\caption{Same as in Table~\ref{tab:bkg_parameters_NGC6791} but for NGC\,6811.}             % title of Table
\centering                         
\begin{tabular}{l l r r}       
\hline\hline
\\[-8pt]         
KIC ID & \multicolumn{1}{c}{$a_\mathrm{meso}$} & \multicolumn{1}{c}{$b_\mathrm{meso}$} & \multicolumn{1}{c}{$\numax$}\\ [1pt]
 & \multicolumn{1}{c}{(ppm)}  & \multicolumn{1}{c}{(\muhz)} & \multicolumn{1}{c}{(\muhz)}\\ [1pt]   
\hline
\\[-8pt]                    
9532903  & $     173.8       ^{+       2.0      }_{-       2.1   } $ & $      29.2       ^{+       0.7      }_{-       0.7   } $ & $      93.9       ^{+       0.6      }_{-       0.4} $\\[1pt]
9534041  & $     150.5       ^{+       0.9      }_{-       0.9   } $ & $      32.4       ^{+       0.5      }_{-       0.4   } $ & $     111.8       ^{+       0.3      }_{-       0.3} $\\[1pt]
9655101  & $     163.6       ^{+       1.2      }_{-       1.2   } $ & $      35.2       ^{+       0.7      }_{-       0.7   } $ & $     101.3       ^{+       0.4      }_{-       0.4} $\\[1pt]
9716090  & $     189.5       ^{+       0.8      }_{-       0.8   } $ & $      30.7       ^{+       0.7      }_{-       0.9   } $ & $     109.6       ^{+       0.2      }_{-       0.2} $\\[1pt]
9716522  & $     244.2       ^{+       3.6      }_{-       4.3   } $ & $      17.3       ^{+       0.4      }_{-       0.5   } $ & $      54.1       ^{+       0.3      }_{-       0.3} $\\[1pt]
9776739  & $     178.0       ^{+       2.6      }_{-       2.3   } $ & $      30.1       ^{+       0.7      }_{-       0.7   } $ & $      94.8       ^{+       0.6      }_{-       0.6} $\\[1pt]    
\hline        
\end{tabular}
\label{tab:bkg_parameters_NGC6811}
\end{table}

\clearpage
\section{Predictions from scaling relations}
\begin{figure*}[htb]
   \centering
   \includegraphics[width=9.0cm]{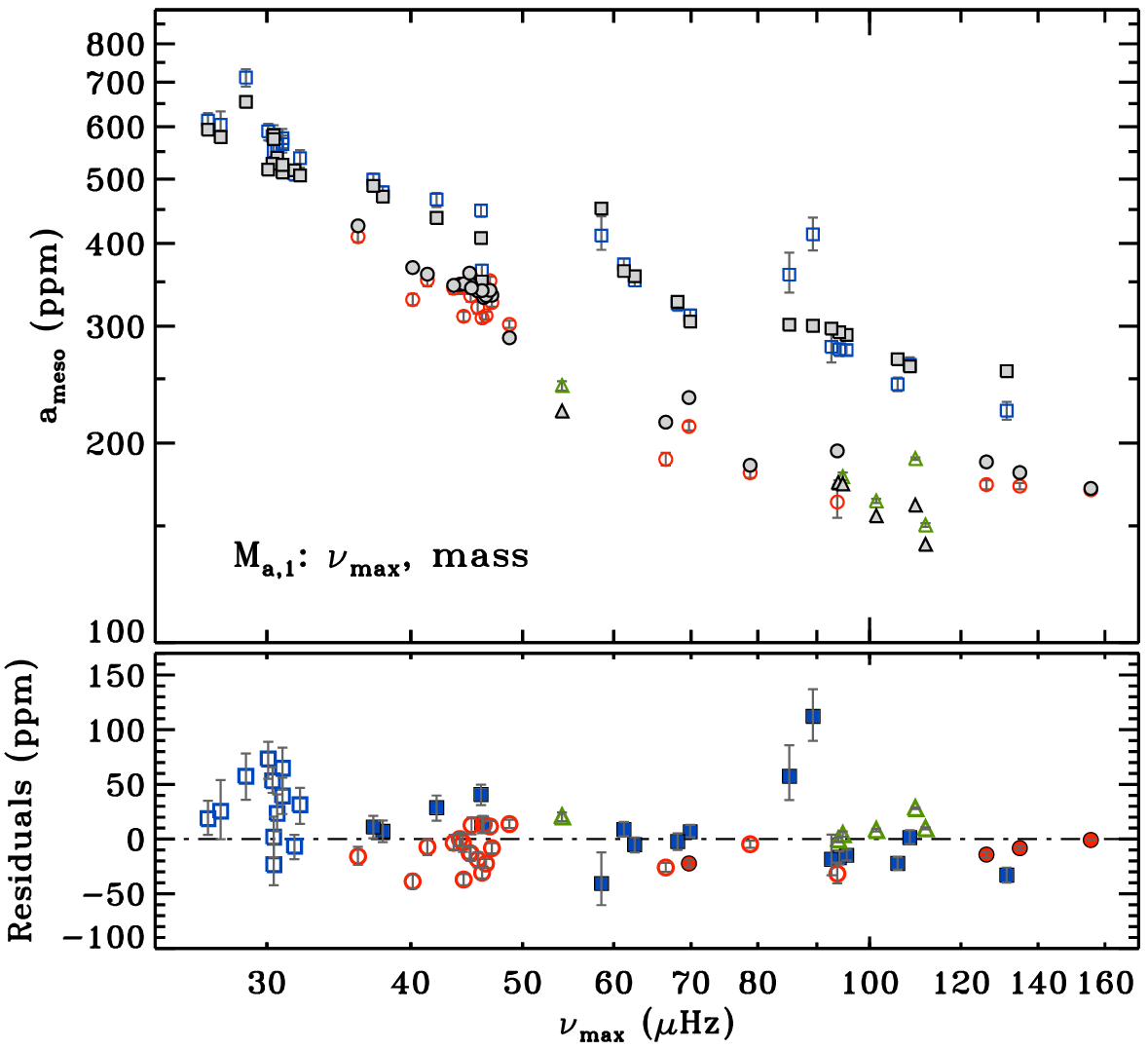}\includegraphics[width=9.0cm]{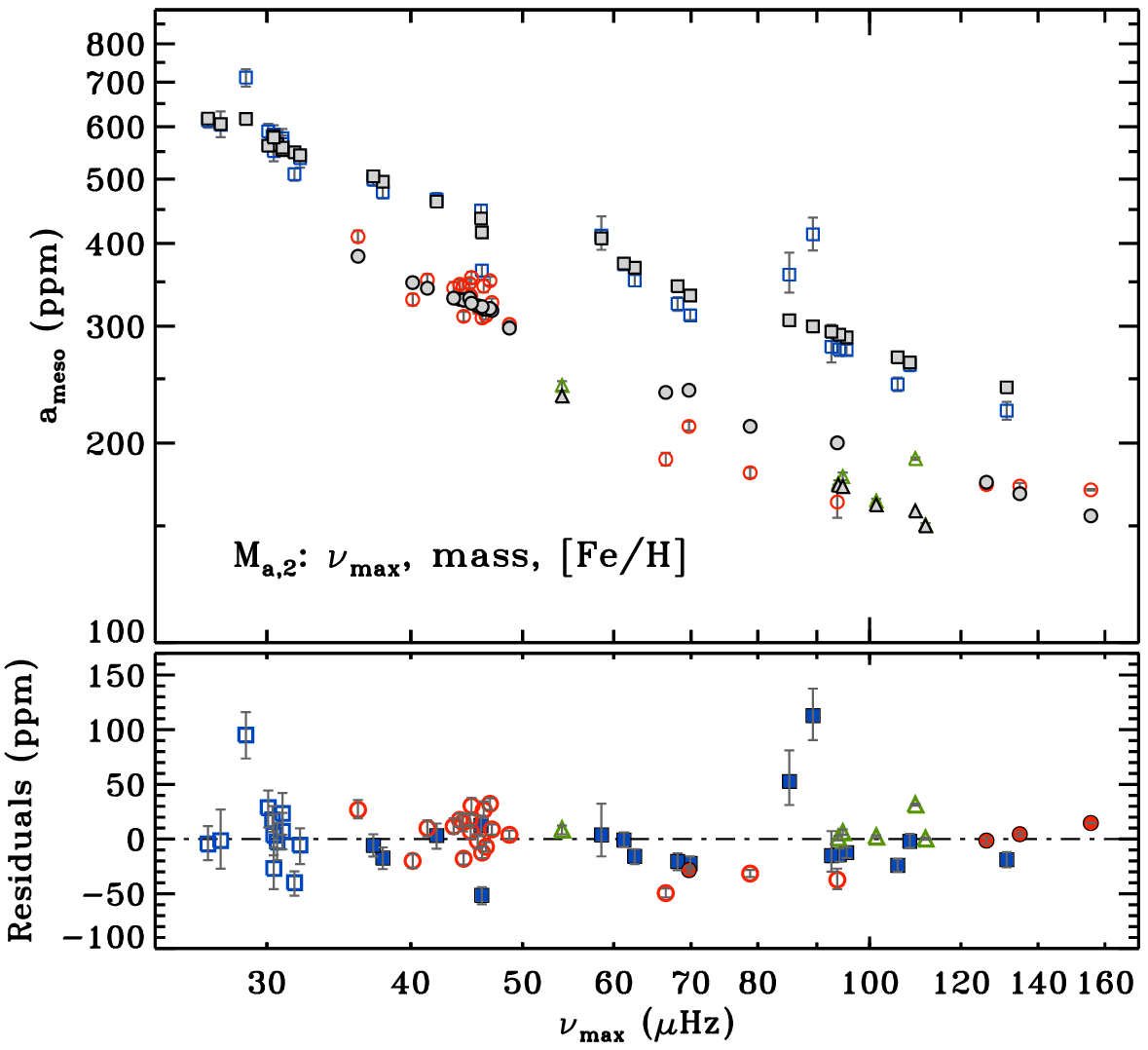}
      \caption{\textit{Top panels}: predicted meso-granulation amplitudes (solid gray symbols) as a function of $\nu_\mathrm{max}$ for NGC\,6791 (squares), NGC\,6819 (circles) and NGC\,6811 (triangles). Median values of the free parameters have been adopted for each scaling relation, as reported in Table~\ref{tab:param_gran}. Observed meso-granulation amplitudes are shown in color with open symbols. The left plot shows the results for the model $\mathcal{M}_{a,1}$, while the right one uses model $\mathcal{M}_{a,2}$. \textit{Bottom panels}: the residuals computed as (Observed-Predicted) meso-granulation amplitudes, with same color and symbol type as in Fig.~\ref{fig:amp_numax}a. Bayesian credible intervals of 68.3\% are overlaid in both panels.}
    \label{fig:fit_amp}
\end{figure*}

\begin{figure*}
   \centering
   \includegraphics[width=9.0cm]{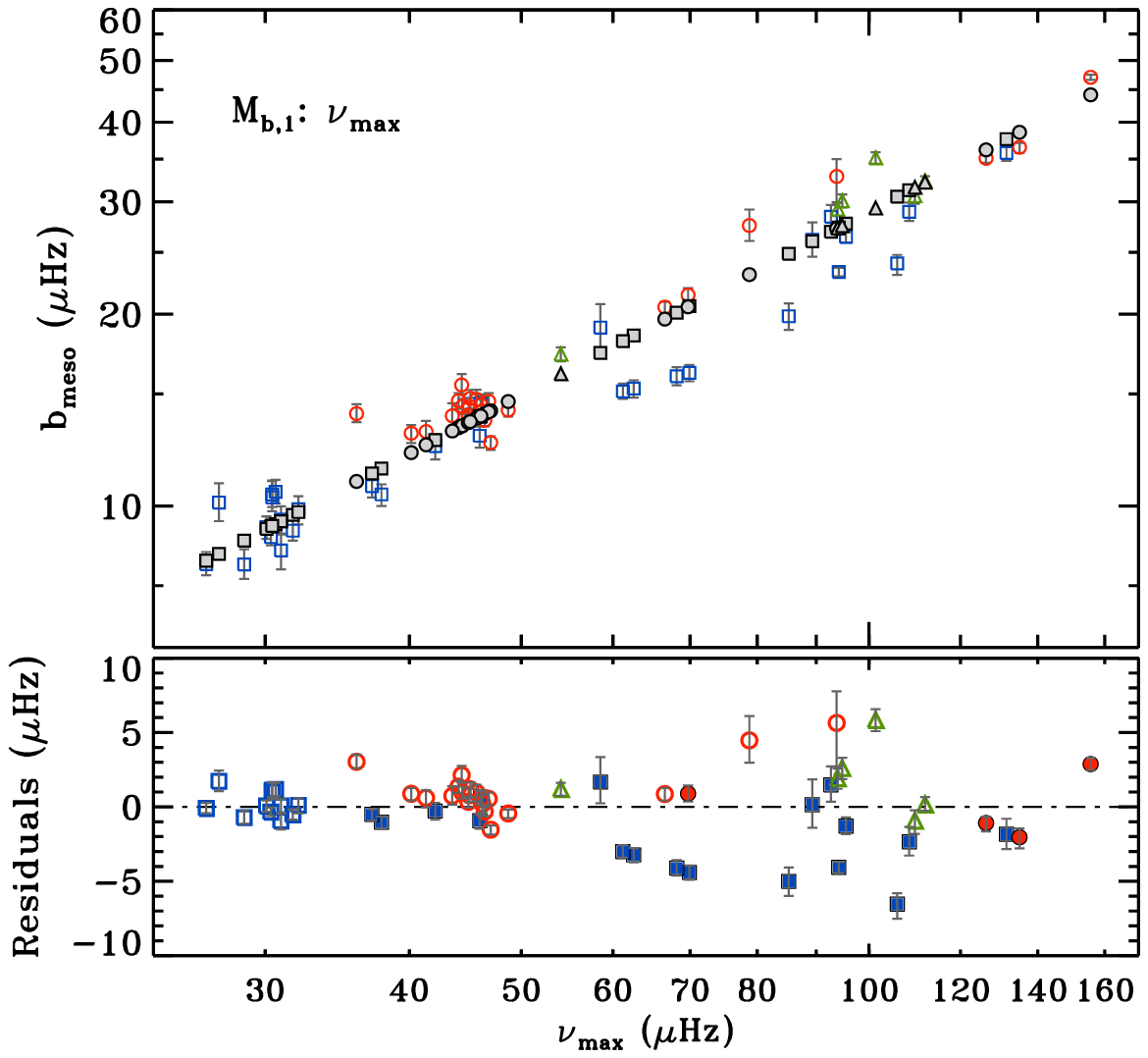}\includegraphics[width=9.0cm]{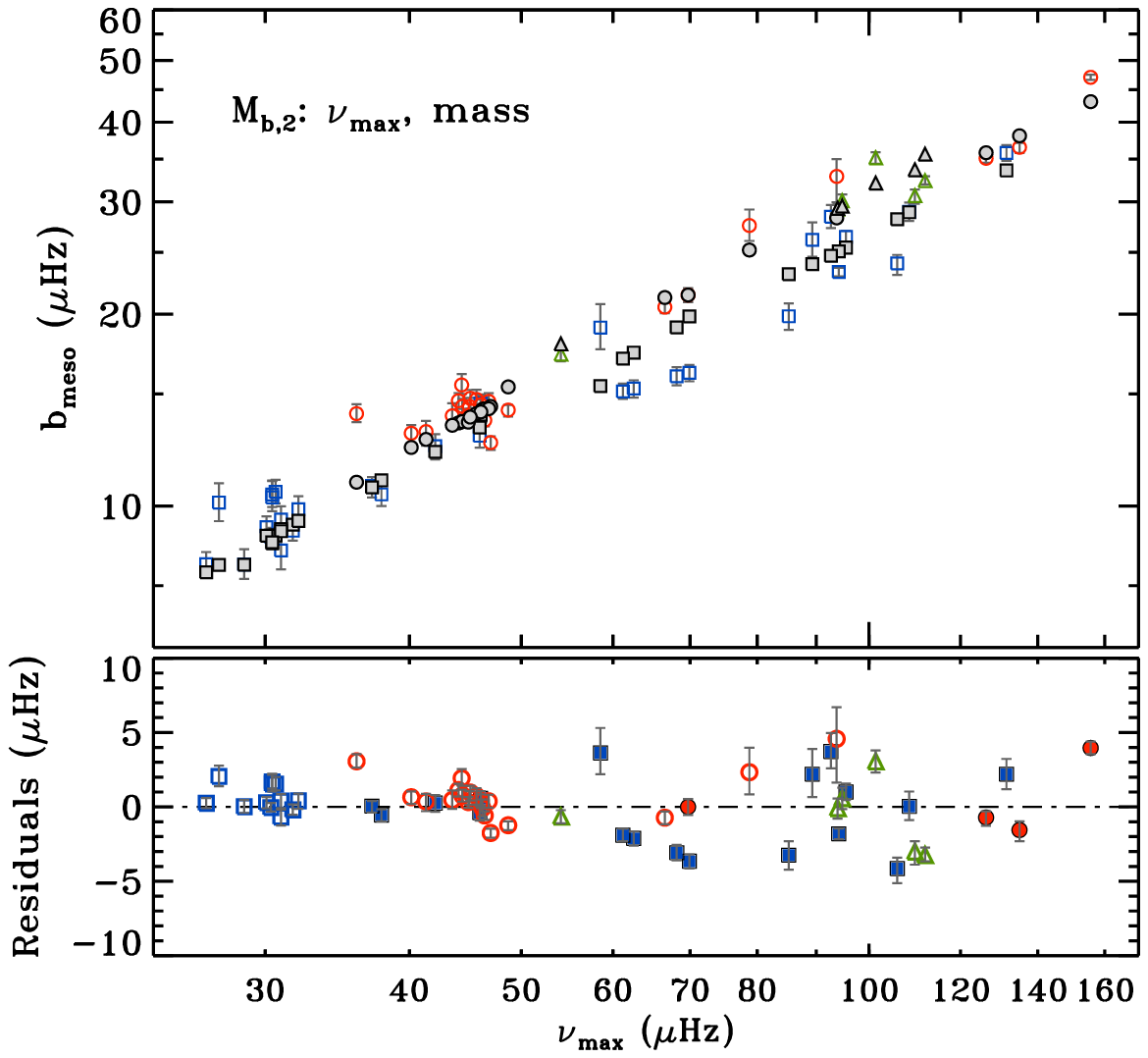}
   \includegraphics[width=9.0cm]{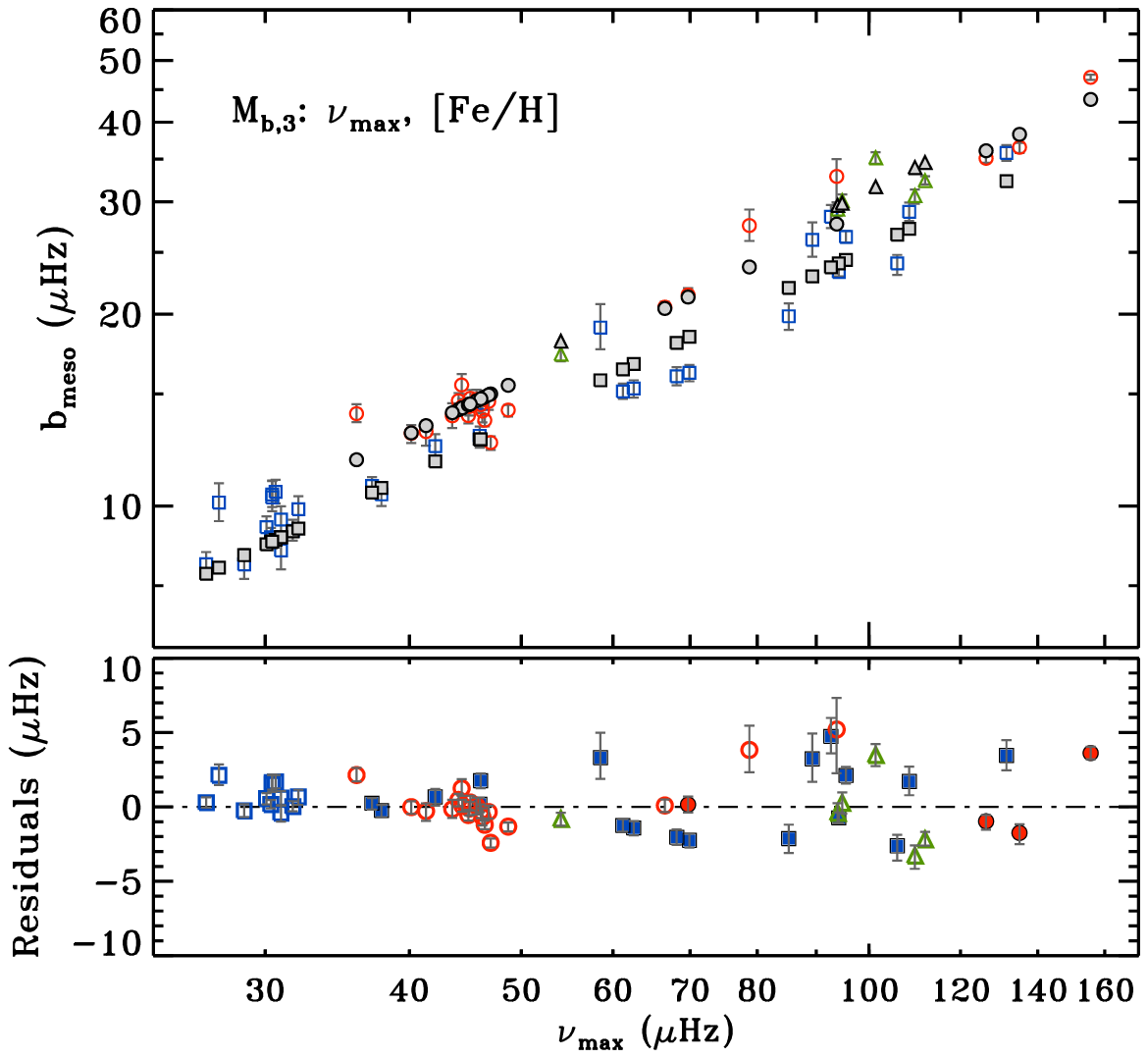}\includegraphics[width=9.0cm]{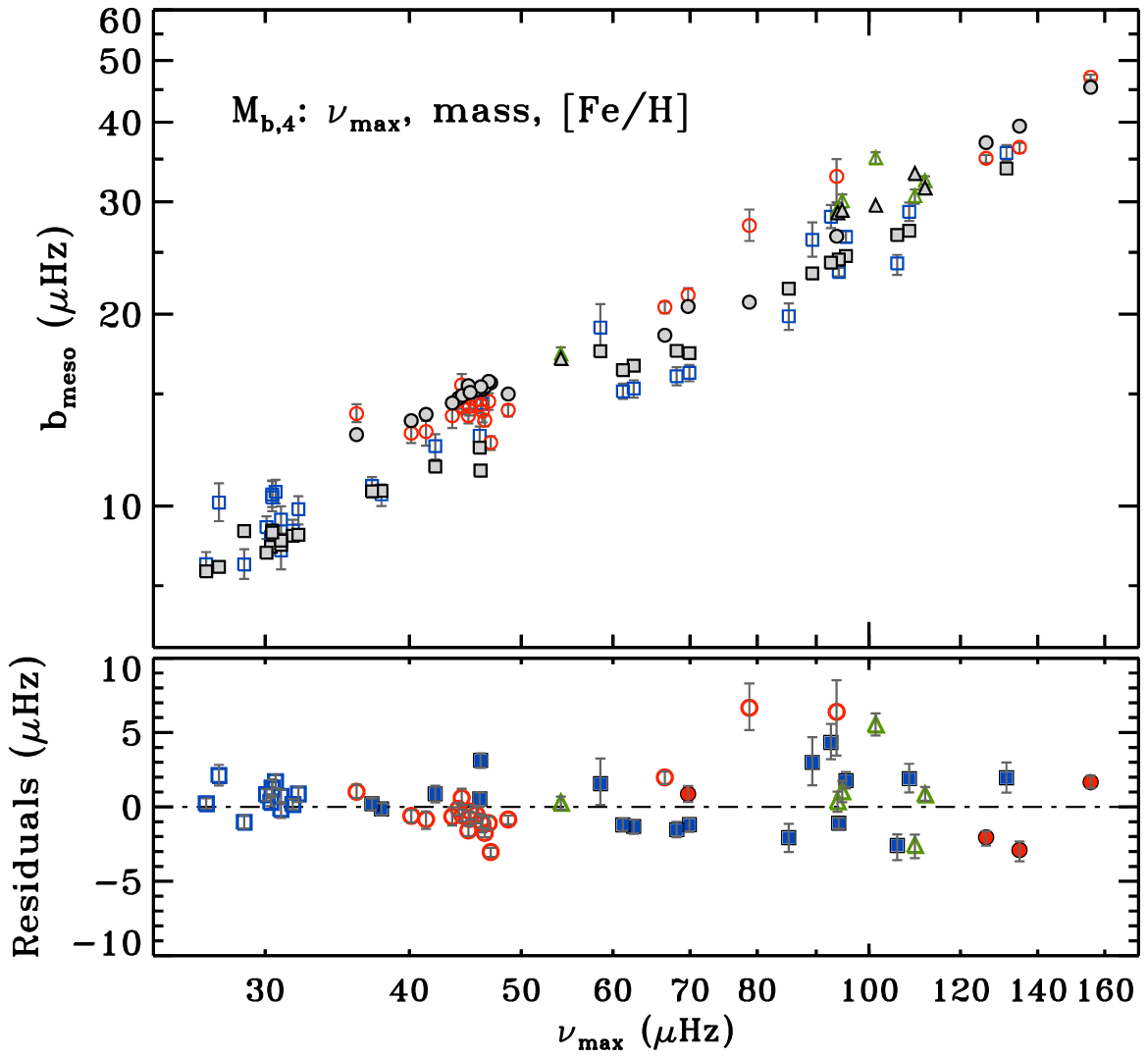}
      \caption{Same as Fig.~\ref{fig:fit_amp} but for the meso-granulation characteristic frequency $b_\mathrm{meso}$. From top left to bottom right we find models $\mathcal{M}_{b,1}$, $\mathcal{M}_{b,2}$, $\mathcal{M}_{b,3}$, $\mathcal{M}_{b,4}$.}
    \label{fig:fit_freq}
\end{figure*}

The resulting predictions of the scaling relations using the estimated parameters listed in Table~\ref{tab:param_gran} are shown in Fig.~\ref{fig:fit_amp} for the models $\mathcal{M}_{a,1}$ and $\mathcal{M}_{a,2}$, and in Fig.~\ref{fig:fit_freq} for the models $\mathcal{M}_{b,1}$, $\mathcal{M}_{b,2}$, $\mathcal{M}_{b,3}$, and $\mathcal{M}_{b,4}$.

\end{document}